\documentclass[longbibliography, %
 reprint,
superscriptaddress,
 amsmath,amssymb,
 aps,
pre
]{revtex4-1}

\usepackage{graphicx}% Include figure files
\usepackage{dcolumn}% Align table columns on decimal point
\usepackage{bm}% bold math
\usepackage{color}
\usepackage{soul}
\usepackage{lipsum} 
\usepackage{hyperref}
\usepackage[english]{babel}
\usepackage{titleref}% add hypertext capabilities
\definecolor{darkgreen}{rgb}{0.0, 0.5, 0.0}

\newcommand{\bl}[1]{\textcolor{black}{#1}}

\begin{document}

\preprint{APS/123-QED}

\title{Bit reset protocols that obey activity-constrained speed limits do not minimize work for a given speed}

\author{Daan Mulder}
 \email{d.mulder@amolf.nl}
\affiliation{AMOLF, Science Park 104, 1098 XG, Amsterdam, The Netherlands}

\author{Thomas E. Ouldridge}
\affiliation{Department of Bioengineering, Imperial College London, London SW7 2AZ, United Kingdom}

\author{Pieter Rein ten Wolde}
\affiliation{AMOLF, Science Park 104, 1098 XG, Amsterdam, The Netherlands}

\date{\today}

\begin{abstract} 
The goal of thermodynamic optimal control theory is to find protocols to change the state of a system from an initial to a desired final distribution, within a finite time, with the least possible expenditure of work. The optimal protocol is closely linked to the intrinsic dynamics of the system at hand. The fact that these dynamics can vary widely from system to system has made a general solution elusive. Recent years have seen great progress by recasting the question in terms of a quantity called total activity, \textit{i.e.} the average number of jumps between states of the system over the course of the operation, rather than the time that the operation is allowed to take. This perspective has allowed for general expressions for the minimal work as a function of the total activity, and the minimal total activity required for a given work. The expression for minimal total activity can be recast as an apparent minimal operation time or speed limit, determined by the average activity rate and work done. A maximal activity rate can be justified by appealing to physical restrictions on the underlying transition rates, but it is unclear whether protocols optimized under a constrained activity actually require the lowest work input for a given operation time  under these restrictions. In the context of bit reset, we show that directly minimizing work for a given operation time under constraints on the rates leads to protocols that require significantly less work to perform the operation than the activity-constrained protocol of the same duration. We show how the resulting protocols for both optimization schemes differ. One reason for the difference between both optimization schemes is the fact that the activity rate is not constant over the course of the protocol: it depends on both the transition rates and the distribution of the bit, both of which change over the course of the copy operation. In the limit of long protocol duration, we find an expression for the difference between the resulting minimal work for both optimization schemes, for a general class of dynamics. The time-constrained approach always outperforms the activity-constrained approach for a given constrained duration, and the difference in work can be arbitrarily large, depending on the boundary conditions and dynamics of the system under consideration. 

\end{abstract}

\maketitle

\section{Introduction}

\noindent Thermodynamics is a powerful framework to describe systems as widely varying as steam engines, computers and enzymes. State functions like entropy and free energy can be used to determine lower bounds on the work required to change the system from one distribution to another. When such a transformation is performed in a finite time, extra work is required. The sophisticated framework of thermodynamic optimal control theory was developed to quantify the minimal required work in the finite-time case \cite{Berry83,crooks07,sivak12,zwanzig2001, peliti2021,sivak2023}. In general, the shorter the duration allowed for the transformation, the higher the extra work, since the system will be forced further out of equilibrium. The result is a trade-off between accuracy, cost and speed that has proven to be relevant in both digital systems \cite{Esposito_2010,Esposito2013, zulkowski2013}, as well as in biology \cite{tu2012,seifert2007, malaguti2021}. 

\begin{figure}
    \centering
    \includegraphics[width=\columnwidth]{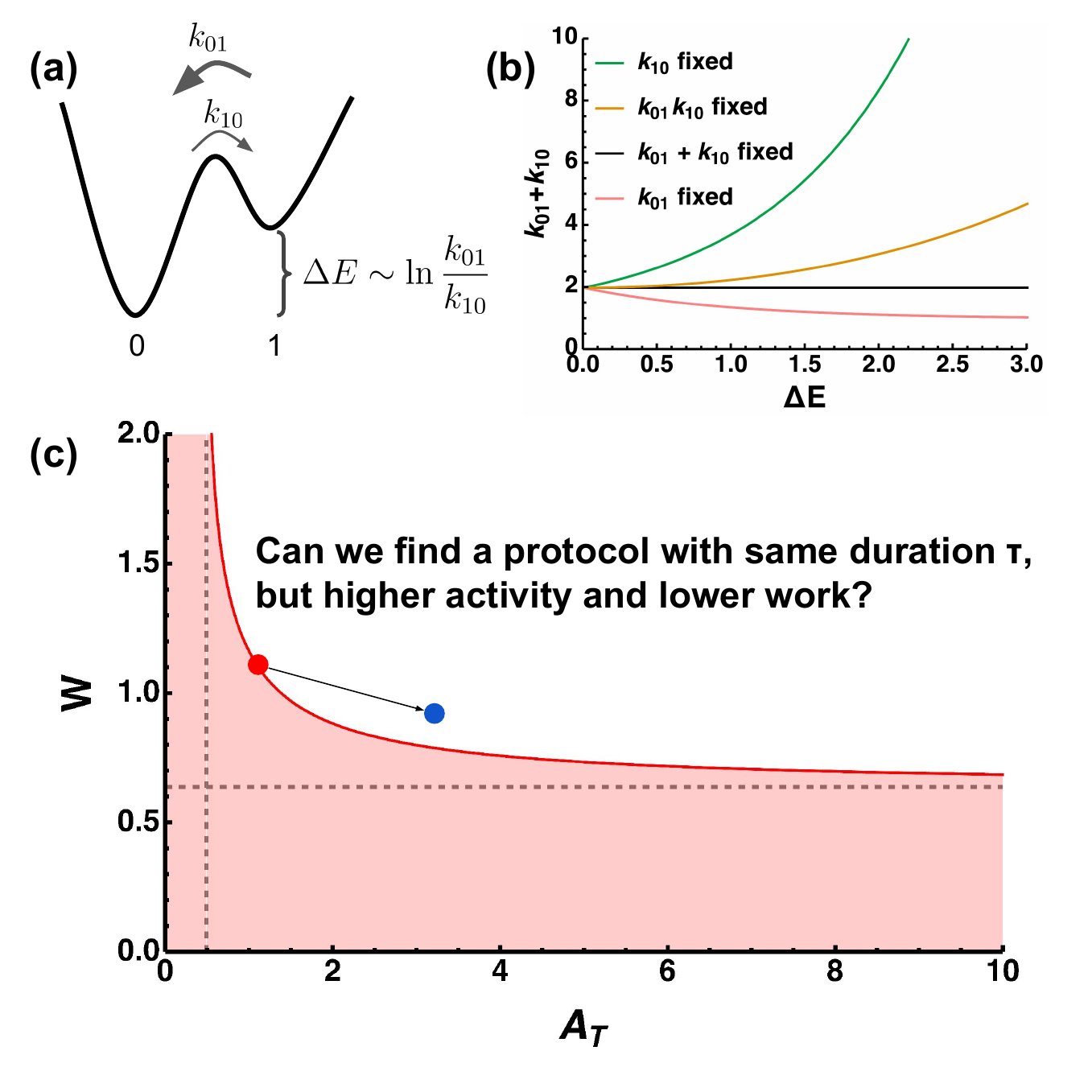}
    \caption{Thermodynamics of bit reset. (a) A bit system with corresponding rates $k_{01}$ and $k_{10}$. The energy difference determines the ratio of the rates through the detailed balance condition. (b) Different relations between the relaxation rate (the sum $k_{01}+k_{10}$) and energy difference are possible, despite the ratio being fixed by the detailed balance condition, e.g. fixing $k_{01}$ (pink) or $k_{10}$ (green), a combination of both which keeps the sum of the two rates equal (black), or their product ($k_{01} \sim e^{\Delta E/ 2 k_B T}$ and $k_{10} \sim e^{-\Delta E/ 2 k_B T} $, orange). All relations are normalized such that $k_{01}=k_{10}=1$ if $\Delta E = 0$ (c) The minimal work to perform a bit reset of an initially symmetric bit with accuracy $s=0.99$ for a given value of $A_T$ equals $W^\text{min}$ given by Eq. (\ref{eq:Wmin}), plotted as a red line. This minimal work leads to a forbidden region under the plot. In the limit $A_T \to \infty$, the work reduces to the quasistatic work (horizontal grid line). The minimal total activity $A_T$ at which the work diverges equals $(s-1/2)$ (vertical grid line). } 
    \label{fig:introduction}
\end{figure}

A prime example of this trade-off arises in finite-time bit reset, see Fig. \ref{fig:introduction}(a) \cite{landauer61}. Apart from important practical considerations like computing efficiency \cite{wong17}, bit reset is of interest since it shows the intimate connection between thermodynamics and information theory \cite{szilard29, Bennett82, ueda09, tenwolde19, mulder2023}. If the operation is performed quasistatically, \textit{i.e.} infinitely slowly, the minimal required work for an unbiased bit is given by the Landauer bound of $k_\text{B} T \ln(2)$, in agreement with experiment \cite{Koski.2014tu, Jun.2014qmn}. For higher speeds, the costs increase \cite{Berut.2012}. Optimal minimal work protocols and minimal finite-time costs have been studied using different techniques like continuous Langevin equations \cite{bechhoefer20} or discrete Markov chains \cite{Esposito2013}. In these optimal protocols, the energy difference between the two states of the bit has a discrete jump at the start of the protocol, and increases gradually after \cite{Esposito2013}. 

When the bit is modeled by a two-state Markov model, the energy difference between the two states of the bit sets the ratio of the rates. While the relative magnitude of the rates is thus fixed, there is a remaining degree of freedom in their absolute magnitude. This degree of freedom can be quantified by different measures like the product of the forward and backward rate of the transition, called the transition width in \cite{maes2017}, or by the sum of the two rates, which we call the relaxation rate. Furthermore, this absolute magnitude can vary as a function of the energy difference.

\bl{The relation} between energy and absolute magnitude of the rates varies from system to system, and must be modeled correctly to determine the optimal protocol for the system at hand. Typical examples of such relations include fixing the relaxation rate itself \cite{Esposito2013, mulder2023}, either of the two individual rates \cite{Seifert_2014,park22}, or by fixing the transition width, so that the forward and backward rate increase respectively decrease by the same factor as the energy difference increases \cite{dechant2022, Remlein.2021}, see Fig. \ref{fig:introduction}(b). For a given constraint on the rates, one might use a Lagrangian approach to calculate the optimal path of the control parameter, and the corresponding minimal work, for a given duration of the protocol. 

There are a few downsides to the Lagrangian optimization scheme that determines the optimal path. First of all, the procedure only works for simple systems with few control parameters, making it hard to extend beyond a two-state system \cite{deweese14, Remlein.2021}. Secondly, even for those simple systems, the procedure does not lead to analytical solutions. Most importantly, the obtained solution is not general, since it depends on the specific relation between energy difference and relaxation time that is imposed. 

Recently, a subtly different approach has led to results for bit reset that are both general and analytical \cite{dechant2022, park22, Vu.2023}. Here, the minimal work is expressed in terms of the average total number of jumps that take place between the states of the system over the course of the protocol, called the total activity $A_T$ \cite{shiraishi18}. For a given value of the total activity, the minimal work can be derived and analytically expressed as a function of the initial distribution, final distribution and total activity. Interestingly, this result does not depend on the relation between energy and the absolute magnitude of the rates. The more jumps are allowed, the closer the system can remain to a quasistatic process, and the lower the minimal work. If only a small number of jumps is allowed, more work is required to ensure these jumps take place in the correct direction. In this sense, the effect of the total activity on the work appears to be similar to that of the protocol duration. Indeed, for a given relaxation rate-energy relation and optimal protocol, one can express the total activity in terms of duration and vice versa \cite{park22, SI}.

The activity-constrained optimization scheme has been applied to study bit reset in finite time \cite{park22, Vu.2023}, and to derive apparent speed limits. Speed limits are expressions that relate the duration $\tau$ to a minimal required work. These expressions can be derived by inverting the expression for the minimal work in terms of the total activity, leading to an expression for a minimal required total activity to perform the operation for a given work. The $\tau$-dependence in the speed limit enters through the definition of the average activity per unit amount of time, defined by \bl{$\langle a \rangle = A_T /\tau$}. For a given work, this definition leads to a bound on $\tau$ that depends on the average activity rate \bl{$\langle a \rangle$}. However, it should be noted that these activity-constrained protocols are not optimized for a given value of protocol duration $\tau$, but rather for a given value of total activity $A_T$. 

\bl{In this paper}, we will show that the optimal protocol that minimizes the work for a given total number of jumps $A_T$, the optimal activity-constrained protocol, is not the same as that optimized for a given duration $\tau$, the optimal time-constrained protocol. For a given protocol duration $\tau$, we will show the existence of a time-constrained protocol (blue point in Fig. \ref{fig:introduction}(c)) that has a lower work than the activity-constrained protocol (red point), yet is suboptimal in terms of the total activity. Under the constraint that seems experimentally most relevant, i.e. time rather than total activity, such a protocol is superior. 

The activity and time constrained optimization schemes lead to different optimal paths, as can be seen by studying quantities like the energy difference, the fluxes and irreversible entropy production rate over the course of the protocol. We will study in which regime this difference is the most significant. How well does an activity constraint serve as a proxy for a time constraint? We find that the difference between the two optimization schemes increases for operations that have a larger change in activity rate over the course of the protocol. We find an expression to quantify this difference in the long-$\tau$ limit, for a class of dynamics of the two-state model. 

The paper is structured as follows. We first introduce the two-state Markov model for bit reset, and the results of \cite{dechant2022,park22}, where the minimal work for a given total activity $A_T$ is derived, as well as the optimal protocol. We will then compare these results to the minimal work and optimal path for a system optimized for a given duration $\tau$, where one of the rates is fixed \cite{Seifert_2014}. \bl{Detailed derivations are provided in the \hyperref[s:appendix]{Appendix}.} The two optimization schemes lead to different optimal paths for the same system: the $\tau$-constrained scheme is optimal for a given duration $\tau$, but suboptimal in terms of $A_T$, and vice versa. By studying different quantities over the course of the protocol, we show how the $\tau$-constrained protocol manages to outperform the $A_T$-constrained protocol for the same $\tau$. We show that in the long-time limit, the ratio of the work for both protocols can be expressed as a function of the ratio of the activity rates over the course of the protocol and validate this result by comparing it to exact solutions. 

\section{Theory}

\subsection{Bit reset and activity-constrained optimization} 

We model our system as a two-state Markov model, where the two states are labeled $0$ and $1$, with transition rates $k_{ij}(t)$ to go from state $j$ to $i$. The master equation of its probability distribution $P_M = \{p_0,p_1\}$ is

\begin{align}
    \dot{p_0} = - \dot{p_1} =  k_{01}(t)p_1(t)-k_{10}(t)p_0(t).
\label{eq:mastereq}
\end{align}

The energy levels and the ratio of the rate constants obey $\Delta E(t) \equiv E_1(t)-E_0(t) =  k_B T \ln(k_{01}/ k_{10})$, so that the equilibrium state obeys the Boltzmann distribution. The energy level serves as the control parameter. In the following, we assume $T= k_B = k_B T =1$. The absolute magnitude of the rates, which we quantify by the relaxation rate $k_{01}+k_{10}$, can depend on $\Delta E$ as well, but this relation is not a priori fixed and can vary from system to system, see Fig. \ref{fig:introduction}(b). The goal of bit reset is to change the system from a symmetric equilibrium distribution ($p_0(0) = p_1(0) = 1/2$, $\Delta E = 0$) to state 0 with accuracy $s > 1/2$, within time $\tau$ so that $p_0(\tau) = s, p_1(\tau) = 1-s$.  In order to reach this desired final distribution, $\Delta E$ needs to be increased over the course of the operation. The required power can be written as $\dot{W} = \sum_i p_i \dot{E_i}$. Integrated from $0$ to $\tau$, the power yields the required work $W$. After the operation, the energy levels are set to their original values. This way, the reset is consistent with a framework where the manipulation of a bit happens through an interaction Hamiltonian acting on the system during the interval $[0,\tau]$ \cite{esposito2017}.

The work $W$ can be separated into a quasistatic part and an irreversible part. The quasistatic, reversible part only depends on the initial and final distribution and equals $\Delta F = \Delta U - T \Delta S$, where $U = \sum_i p_i E_i$ is the internal energy, and $S = - k_B \sum_i p_i \log(p_i)$ is the entropy. In the case of an initially symmetric bit, $\Delta U = 0$ since $U(0) = U(\tau) = 0$ as the energy levels are initially equal and are set to the initial value after the operation. Next to the quasistatic part there will also be an irreversible part $W_\text{irr}$, due to the fact that the system needs to be brought out of equilibrium to achieve the reset in a finite time. Contrary to the quasistatic costs, these depend on the path that $\Delta E(t)$ takes over the interval $[0,\tau]$ in order to bring the system to the final distribution. 

The central result of \cite{dechant2022,park22} is to find an analytical result for the minimal work for a given total activity $A_T$, defined as the average total number of jumps that happens over the course of the operation,

\begin{align}
  A_T  = \int_0^\tau dt \hspace{3mm} k_{01}(t)p_1(t) + k_{10}(t)p_0(t).
\label{eq:activity}
\end{align}
The integrand, called traffic in \cite{maes2008}, will be referred to as the activity rate \bl{$a(t)$}. Given a total activity $A_T$, the minimal work required to reset an initially symmetrically distributed bit with accuracy $s$, so that the bit is in state $0$ with probability $s$ after the reset, equals

\begin{align}
 W^\text{min}|_{A_T} = \Delta S + (s-1/2) \ln \left( \frac{1+\frac{s-1/2}{A_T}}{1-\frac{s-1/2}{A_T}} \right),
\label{eq:Wmin}
\end{align}
where the first term is the quasistatic, reversible cost due to the Landauer bound. The change in Shannon entropy, $\Delta S = \ln(2) + s \ln s + (1-s)\ln(1-s)$, equals $\ln(2)$ when $s=1$. The second term is the irreversible cost $W_\text{irr}^\text{min}|_{A_T}$. We show the minimal work $W^\text{min}|_{A_T}$ in Fig. \ref{fig:introduction}(c) for a reset with accuracy $s=0.99$. The irreversible work decreases to $0$ with increasing $A_T$, so that only the quasistatic cost remains in the high $A_T$-limit. Surprisingly, the minimal required work only depends on the initial and final distribution of the system and the total activity $A_T$, regardless of the relation between energy and relaxation rate that the system obeys.

For a given value of $A_T$, the optimal protocol is achieved by varying the energy difference so that 

\begin{align}
  K = \frac{k_{01}(t) p_1(t)}{k_{10}(t) p_0(t)}
\label{eq:KA}
\end{align}
is a constant of motion. \bl{Both the constant of motion and the corresponding minimal work are derived in \cite{park22}, see also Appendix \ref{ss:Kconstant}.} This constant of motion implies that the generalized force $\ln(K)$ is constant over the course of the operation. The work is then given by Eq. (\ref{eq:Wmin}). It follows from Eq. (\ref{eq:Wmin}) that the minimal required total activity is $A_{T,\text{min}} = (s-1/2)$. This is the scenario where no jumps from $0$ to $1$ are allowed, so all jumps increase $p_0$. Then, only $(s-1/2)$ jumps are needed on average to make the transformation from $p_0(0) = 1/2$ to $p_0(\tau) = s$. In this limit, the ratio $k_{01}/k_{10}$ and hence the generalized force and work diverge. 

Crucially, we need to fix the relationship between the energy difference $\Delta E$ and the relaxation rate, set by the magnitudes of the rates, in order to obtain a unique activity-constrained optimal protocol for a given total activity. While any protocol that obeys Eq. (\ref{eq:KA}) is optimal, in the sense that it minimizes the work for a given $A_T$, this protocol is not unique in the absence of another constraint on the magnitudes of the rates. There exists an ensemble of protocols that all minimize the work for a given $A_T$, obeying Eq. (\ref{eq:KA}), and any pair of protocols in this ensemble can be mapped onto one another by multiplying both the forward and backward rate by the same time-dependent prefactor $\alpha(t)$. For example, an optimal protocol that minimizes the work for a given $A_T$ can be mapped onto another protocol with the same work, $A_T$ and, indeed, the same duration $\tau$, if both the forward and backward rate are increased by a certain factor in the first part of the protocol and decreased by another factor in the second part. This change corresponds to speeding up the movie of the physical process in the first half and slowing it down in the second half. In general, however, the relaxation rate of the system is bounded by the physics of the system, which, in our framework, manifests itself as a relationship between the rates and $\Delta E$ (see Fig. \ref{fig:introduction}(b)). With this other constraint, there is a unique activity-constrained optimal protocol that minimizes the work for a given $A_T$. This protocol has a specific duration $\tau$; the larger $A_T$, the larger $\tau$, and vice versa. 

The theoretical framework presented in \cite{dechant2022, park22, Vu.2023} is more general than the specific implementation presented here. For example, Eq. (\ref{eq:Wmin}) is the solution for the work in the case of a reset of a bit, starting from a symmetric initial distribution. The general solution for the asymmetric case can be found in \cite{park22}, \bl{see also Appendix \ref{ss:Kconstant}}. When the system has a more complex topology than a two-state system, the distance measure $(s-1/2)$ that occurs in Eq. (\ref{eq:Wmin}) needs to be replaced by a more general Wasserstein distance \cite{Vu.2023, dechant2022}, which connects optimal transport theory with thermodynamics in the case of both discrete \cite{hasegawa2021} and continuous state space \cite{aurell11,ito21}. It should be noted that in \cite{Vu.2023}, the results are not presented for the total activity but rather a quantity called kinetic cost $M_T = \int_0^\tau m(t) dt$, where $m(t)$ is the so-called dynamical state mobility $m(t)$, defined as
\begin{align}
    m(t) = \frac{k_{01}(t)p_1(t) - k_{10}(t)p_0(t)}{\ln \left(\frac{k_{01}(t)p_1(t)}{k_{10}(t)p_0(t)} \right)}.
\end{align}
The minimal irreversible work for a given value of $M_T$ takes on an especially simple form, with $W_\text{irr}^\text{min}|_{M_T} = (s-1/2)^2/ M_T$. Despite the difference in appearance compared to the irreversible work for the activity constrained protocol (see Eq. (\ref{eq:Wmin})), both optimization schemes lead to the same optimal protocol, which has a constant generalized force like Eq. (\ref{eq:KA}), so that all conclusions regarding activity-constrained protocols carry over to dynamical-state-mobility constrained protocols, as well as the generalized activity metrics considered in \cite{ito2024}. 

The equation for the minimal work (Eq. (\ref{eq:Wmin})) can also be cast in the form of a speed limit as

\begin{align}
    \tau \geq \frac{(s-\frac{1}{2})}{\bl{\langle a \rangle }} \text{coth} \left( \frac{W_\text{irr}}{2(s-\frac{1}{2})}\right)
    \label{eq:speedlimit}
\end{align}
where $W_\text{irr}$ is the irreversible part of the work and \bl{$\langle a \rangle = A_T / \tau$} is the activity rate, averaged over the course of the protocol. The optimal activity-constrained protocol saturates the inequality. No protocol can perform a reset of accuracy $s$ with a lower $\tau$, unless either $W_\text{irr}$ or \bl{$\langle a \rangle$} is increased. The average activity rate \bl{$\langle a \rangle$} can be increased by performing the bit reset using a different system with higher rates, but for a given system with a specified relation between $\Delta E$ and the rates, then, for a given $\tau$, the bound is only saturated for one protocol,  the optimal protocol, which has a specific value of \bl{$\langle a \rangle$}. However, for
that system and that $\tau$, other protocols do exist, which will, in general, have a different $\langle a \rangle$. These protocols will not saturate the bound of Eq. \ref{eq:speedlimit}. But could they have a lower work? To be more precise: can, for a given system with a specific relation between the rates and $\Delta E$, the reset be performed with the same accuracy and within the same time $\tau$, but with a lower work because \bl{$\langle a \rangle$} and hence $A_T = \bl{\langle a \rangle} \tau$ is higher?  This hypothetical protocol would correspond to the blue point in Fig. \ref{fig:introduction}(c). Note that the blue point is not at the bound, since, for the same relaxation rate-energy relation, different points at the bound correspond to different values of $\tau$, whereas the blue and red point have the same value of $\tau$. 

We compare the activity-constrained optimization scheme to optimal protocols that optimize the work directly for a certain value of $\tau$ ($\tau$-constrained optimization scheme). Crucially, optimizing for a certain value of $\tau$ requires that a relation between relaxation rate and energy difference is specified. In this case, we fix $k_{01} = 1$: the rate of going to the desired $0$ state stays constant over the course of the protocol. Using a $\tau$-constrained optimization scheme, we will find a protocol that, for a given value of $\tau$, has a higher value of $A_T$ and a lower work than the activity-constrained protocol, corresponding to the blue point in Fig. \ref{fig:introduction} (c). This protocol will have a different constant of motion than that of Eq. (\ref{eq:KA}). Hence, it is not on the optimal line for a given value of $A_T$, and does not saturate the speed limit of Eq. (\ref{eq:speedlimit}). However, it does outperform the activity-constrained protocol for the given value of $\tau$. 

\subsection{Optimization for a given duration} 

Our goal is to minimize the work for a given duration of the protocol $\tau$. For a given value of the initial and final distribution and energy levels, the quasistatic, reversible part of the work cannot be optimized since it is fixed by the boundary conditions. Instead, we focus on the irreversible work, defined as \cite{peliti2021}

\begin{align}
    W_\text{irr} = \int_0^\tau dt \dot{p}_0 \ln \left( \frac{k_{01}p_1}{k_{10}p_0} \right) = \int_0^\tau dt \dot{p}_0 \left(\Delta E -\ln \left( \frac{p_1}{p_0} \right) \right).
    \label{eq:Wirr}
\end{align}
All quantities under the integral sign are time-dependent. Before minimizing Eq. (\ref{eq:Wirr}), we note that in the limit of diverging energy difference $\Delta E$, the ratio between the rates diverges, so that $k_{10} \approx 0$. Since $k_{01}=1$, Eq. \ref{eq:mastereq} reduces to $\dot{p_1} = - p_1(t)$, implying $p_1(t)$ decays exponentially. This is the fastest $p_0$ can increase. In order to reach the desired accuracy, a minimal value $\tau_\text{min} = - \log(2(1-s))$ is required. For this value of $\tau$, $W_\text{irr}$ diverges. 

The optimal path of $\Delta E(t)$ on the interval $[0,\tau]$ can be found by interpreting $W_\text{irr}$ as an action, so that the integrand is a Lagrangian $\mathcal{L} = \dot{p}_0 \ln(k_{01}p_1/k_{10} p_0)$ \cite{Esposito_2010}. Using the fact that $p_1 = 1-p_0$, as well as the master equation Eq. (\ref{eq:mastereq}) and the fact that $k_{10} = 1$, the Lagrangian can be written solely in terms of $p_0$, $\dot{p_0}$ and constants, as

\bl{\begin{align}
    \mathcal{L} = \dot{p_0} \ln \left( \frac{p_1}{k_{10}p_0} \right) =  \dot{p_0} \ln \left( \frac{1-p_0}{1-p_0-\dot{p}_0} \right).
\end{align}}

\noindent See \bl{Appendix \ref{ss:tauk01}} for details. For such a Lagrangian, with no explicit time dependence, the action is minimized for a path with the constant of motion $\kappa = \dot{p_0}\frac{\partial \mathcal{L}}{\partial \dot{p_0}}- \mathcal{L} $, which leads to 

\begin{align}
    \kappa = \frac{\dot{p_0}^2}{1-p_0 - \dot{p}_0}.
    \label{eq:kappa}
\end{align}
Comparing $\kappa$ (Eq. (\ref{eq:kappa})) for the time-constrained optimization scheme with $K = (1-p_0)/(1-p_0 -\dot{p_0})$ for the activity-constrained protocol (see Eq. (\ref{eq:KA})), using the fact that $k_{01}=1$ and using Eq. (\ref{eq:mastereq}) to substitute $k_{10}p_0 = 1-p_0 - \dot{p_0}$), shows that $K$ and $\kappa$ can only both be constant if $1-p_0 \propto \dot{p_0}^2$. However, this relation does not hold for the optimal solution of the $A_T$-constrained protocol, \bl{as derived in Appendix \ref{ss:Ak01}}, giving a first hint that the two protocols are different. 

The minimal irreversible work for the $\tau$-constrained optimization scheme can be expressed in terms of $\kappa$, $p_0(0)$, $p_0(\tau)$, see \cite{Esposito2013} \bl{and Appendix \ref{ss:tauk01}}. The constant of motion $\kappa$ is related to the boundary conditions $p_0(0)$ and $p_0(\tau)$, and $\tau$ via a transcendental equation, which can be numerically solved to calculate the minimal irreversible work for a given value of $\tau$, denoted $W_\text{irr}^\text{min}|_\tau$. The corresponding activity can be calculated by numerically performing the integral in Eq. (\ref{eq:activity}). Like in the case of the activity-constrained optimization scheme, $A_T$ and $\tau$ are related through a monotonic relation.

\begin{figure*}
    \centering
    \includegraphics[width=\textwidth]{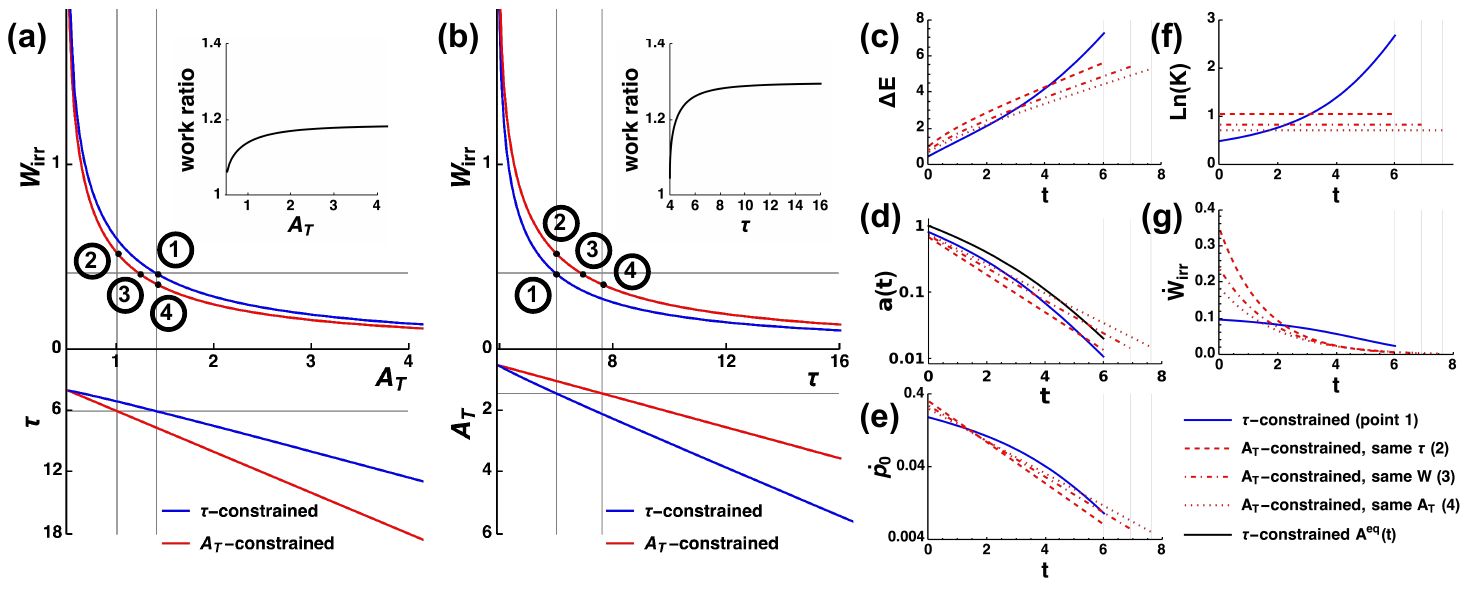}
    \caption{The protocol that minimizes the work for a given duration $\tau$ does not lead to the minimal work for the corresponding value of $A_T$, and vice versa: the activity-constrained protocol is not optimal for a given duration $\tau$. The two optimization schemes lead to markedly different optimal protocols. We study a reset that takes the bit from a symmetric initial state to state $0$ with accuracy $s=0.99$, and dynamics with a fixed forward rate $k_{01}=1$. Panel (a) shows, as a function of the total activty $A_T$, the minimal irreversible work for that $A_T$, $W_\text{irr}^\text{min}|_{A_T}$ (in red), as well as the irreversible work for a protocol that minimizes the work for a given, yet different duration $\tau$, $W_\text{irr}^\text{min}|_{\tau}$ (in blue), yet with the same total activity. 
    Below the x-axis, we plot the corresponding value of $\tau$ for both protocols.
    The protocol optimized for a given value of $\tau$ has a higher work than the one optimized for $A_T$. However, for the same value of $A_T$, it performs the operation in a shorter time $\tau$. The inset shows the ratio between the two required amounts of work, $W_\text{irr}^\text{min}|_{\tau}/W_\text{irr}^\text{min}|_{A_T}$.  Conversely, in (b), we show  $W_\text{irr}^\text{min}|_{A_T}$ (again in red) and $W_\text{irr}^\text{min}|_{\tau}$ (again in blue) as a function of $\tau$, with the corresponding value of $A_T$ plotted below the $x$-axis. Note that this is essentially the same plot as (a), but with the axes for $\tau$ and $A_T$ switched. For a given value of $\tau$, the protocol optimized for a given value of $A_T$ has a higher work than the one optimized for $\tau$, but performs the operation with a \bl{lower total activity $A_T$}. The ratio $W_\text{irr}^\text{min}|_{A_T}/W_\text{irr}^\text{min}|_{\tau}$ is shown as an inset. The points 1-4 correspond to the same protocols in panels (a) and (b), respectively. Panels (c)-(g) show different performance-related quantities as a function of time $t$, over the course of the protocol. We compare the $\tau$-constrained protocol corresponding to point 1 in panel (a) and (b) ($\tau=6$, plotted in blue) with $A_T$-constrained protocols that have the same duration $\tau$ (point 2 in panel (a) and (b), dashed red lines), the same work (point 3, dot-dashed red lines) and the same total activity (point 4, dotted red lines). The endpoints of the protocols are marked with grey vertical lines. \bl{The behaviour of the protocols is discussed in the main text. (c) Energy difference $\Delta E$. (d) Activity rate \bl{$a(t)$} on a logarithmic scale. The black line gives the equilibrium activity rate (defined in Eq. (\ref{eq:eqA})) for the $\tau$-constrained protocol. (e) Flux $\dot{p}_0(t)$ on a logarithmic scale. (f) Generalized force $\ln (K) $, with $K$ given as in Eq. (\ref{eq:KA}). (g) Irreversible entropy production $\dot{W}_\text{irr}$. The area under the blue line and dot-dashed red line is the same since both have the same work.}} 
    \label{fig:worktauA}
\end{figure*}

\section{Results}

\subsection{Comparing the two optimization schemes} The optimal protocol that minimizes the work for a given total activity differs from the protocol that minimizes the work for a given duration. This fact can be seen in Fig. \ref{fig:worktauA}, which shows the minimal work as a function of the total activity $A_T$ (panel (a)), and as a function of the protocol duration $\tau$ (panel (b)), for both the activity-constrained (red lines) and $\tau$-constrained protocol (blue lines). The bit reset has an accuracy of $s=0.99$. For both optimization schemes, each value of the total activity $A_T$ corresponds to a certain value of $\tau$ and vice versa. These relations are shown in the same panels as those of the work, beneath the x-axis. 

\bl{For a given activity $A_T$,}$W_\text{irr}^\text{min}|_{A_T}$ is always lower than $W_\text{irr}^\text{min}|_\tau$. However, the $A_T$-optimal protocol takes longer than the $\tau$-optimal protocol. Comparing for the same value of $\tau$ shows that $W_\text{irr}^\text{min}|_\tau$ is lower than $W_\text{irr}^\text{min}|_{A_T}$, as can be seen explicitly in panel (b), where the same quantities are plotted as a function of $\tau$. To elucidate the difference, we marked point $1$ in panel (a) and (b), which is optimized for duration $\tau = 6$ \bl{(in units of $1/k_{10}=1$)}, with three protocols optimized for $A_T$: point 2, which has the same duration $\tau$ as point 1, a lower activity $A_T$ and a higher work $W_\text{irr}$; point 3, which has the same work, a smaller activity and a longer duration \bl{($\tau \approx 6.9$)}; and point 4 \bl{($\tau \approx 7.6)$}, which has the same total activity, a longer duration and a lower work. The results show that the activity-constrained optimization scheme can indeed be outperformed for a given duration $\tau$, by a protocol directly optimized for duration. 

The two optimization schemes never lead to the same protocol, except in the limit of diverging work, where both $A_T$ and $\tau$ take their minimal value. In all other cases, including the limit of long duration $\tau$ and large activity $A_T$, the optimization schemes lead to different values of the irreversible work. In the insets of Fig. \ref{fig:worktauA}(a,b), we plot the ratio of the irreversible work for both optimization schemes, dividing the the largest of the two values by the smallest, \textit{i.e.} in panel (a) the ratio is $W_\text{irr}^\text{min}|_{\tau}/W_\text{irr}^\text{min}|_{A_T}$, and in (b) the ratio is $W_\text{irr}^\text{min}|_{A_T}/W_\text{irr}^\text{min}|_{\tau}$. This ratio is an increasing function which starts at $1$ for $A_T = A_{T,\text{min}}$ and $\tau = \tau_\text{min}$, but quickly increases and then approaches a maximal value. Notably, the difference can be more than 20\%. Before determining which quantities influence the size of this difference in the next section, we compare the optimal protocols in detail.  

How do the two optimal protocols differ, in terms of the path of the energy difference $\Delta E(t)$ over the course of the protocol, as well as the resulting fluxes and irreversible entropy production? To answer this question, we show the energy difference $\Delta E (t)$, the activity rate \bl{$a(t)$} (the integrand of Eq. (\ref{eq:activity})), the flux $\dot{p_0}(t)$ (see Eq. (\ref{eq:mastereq})), the generalized force $\ln(K(t))$ (see Eq. (\ref{eq:KA})) and the entropy production $\dot{W}_\text{irr}(t)$ in Fig. \ref{fig:worktauA}, panels (c)-(g).  We compare the protocol that corresponds to point $1$ in panels (a) and (b), with the three protocols optimized for $A_T$ corresponding to point 2 (same duration, dashed lines), point 3 (same work, dot-dashed lines) and point 4 (same total activity, dotted lines).  Analytical expressions for the quantities shown in the case of $A_T$-constrained optimization can be found in Appendix \ref{ss:Ak01}.

There are a few similarities between the protocols resulting from the two optimization schemes. The energy difference between the two states $\Delta E(t)$ (panel (c) of Fig. \ref{fig:worktauA}) has to increase over time in order to perform the reset operation. Furthermore, due to the constraint on the rates where the forward rate $k_{01}=1$ is kept constant, an increasing value of $\Delta E(t)$ implies that the backward rate $k_{10}(t)$ decreases over time. The decrease of the rate leads to a decrease in \bl{$a(t)$} over time as well, regardless of whether the protocol is optimized for $\tau$ or $A_T$, as shown in panel (d). Similarly, we see that the flux $\dot{p}_0(t)$ and the irreversible entropy production $\dot{W}_\text{irr}$ decrease (panel (e) and panel (g)) over the course of the protocol. For the activity-constrained protocols, the activity rate \bl{$a(t)$}, flux $\dot{p_0}(t)$ and irreversible entropy production $\dot{W}_\text{irr}$ decay exponentially, leading to straight lines on the logarithmic plots of panel (d) and (e). 

Despite their similarities, the protocols optimized for $A_T$ have very different properties from the protocols optimized for $\tau$. The $A_T$-constrained protocols have a larger energy difference $\Delta E$ at the start of the protocol, but a weaker increase of $\Delta E$ over time, \bl{so that the final value of $\Delta E$ is smaller} (Fig. \ref{fig:worktauA}(c)) We explain \bl{the difference between the two optimization schemes in the next section. We can, however, explain the difference in the other quantities from the difference in the path of $\Delta E(t)$}. At the start of the protocol, using the fact that $p_0(t) \approx p_1(t) \approx 1/2$, and $k_{01}=1$, the activity rate can be approximated as \bl{$a(t) \approx 1/2(1+k_{10})$}. Since an increased $\Delta E$ implies a decreased backward rate $k_{10}$, the $A_T$-constrained protocols have a lower activity at the start. \bl{Over the course of the protocol, the activity rate of the $\tau$-based protocol decreases more strongly, so its final activity is lower than the $A_T$-based protocols.} 

\bl{The flux $\dot{p}_0(t)$ can be approximated at the start of the protocol as $p_0(t) \approx 1/2(1-k_{10})$, which is initially larger for the $A_T$-constrained protocols given the initial values of $\Delta E$.} Panel (f) shows the generalized force, which is equal to $\ln(K)$, and hence constant for the $A_T$-constrained protocols (see Eq. (\ref{eq:KA})). Since the initial energy difference is larger for the $A_T$-constrained protocol, it also has a higher initial generalized force than the $\tau$-constrained protocol. The price to pay for the quick start of the $A_T$-constrained protocol is a larger irreversible entropy production initially (panel (g)), which is the product of the flux and the generalized force (see Eq. (\ref{eq:Wirr})). The $\tau$-constrained protocol catches up in the second half of the protocol: $\Delta E$ surpasses that of the $A_T$-constrained protocols, the activity rate is lower, the flux becomes higher, as well as the generalized force and irreversible entropy production. \bl{Note that the irreversible entropy production for the $\tau$-constrained protocol is relatively constant compared to the $A_T$-constrained protocol. For larger values of $\tau$, the change in irreversible entropy production gets smaller and smaller, in line with the fact that for slow-varying $\tau$-constrained optimal protocols, the irreversible entropy production is constant \cite{sivak12}.}

\begin{figure}[t]
    \centering
    \includegraphics[width=\columnwidth,trim = 0 0.5cm 0 0, clip]{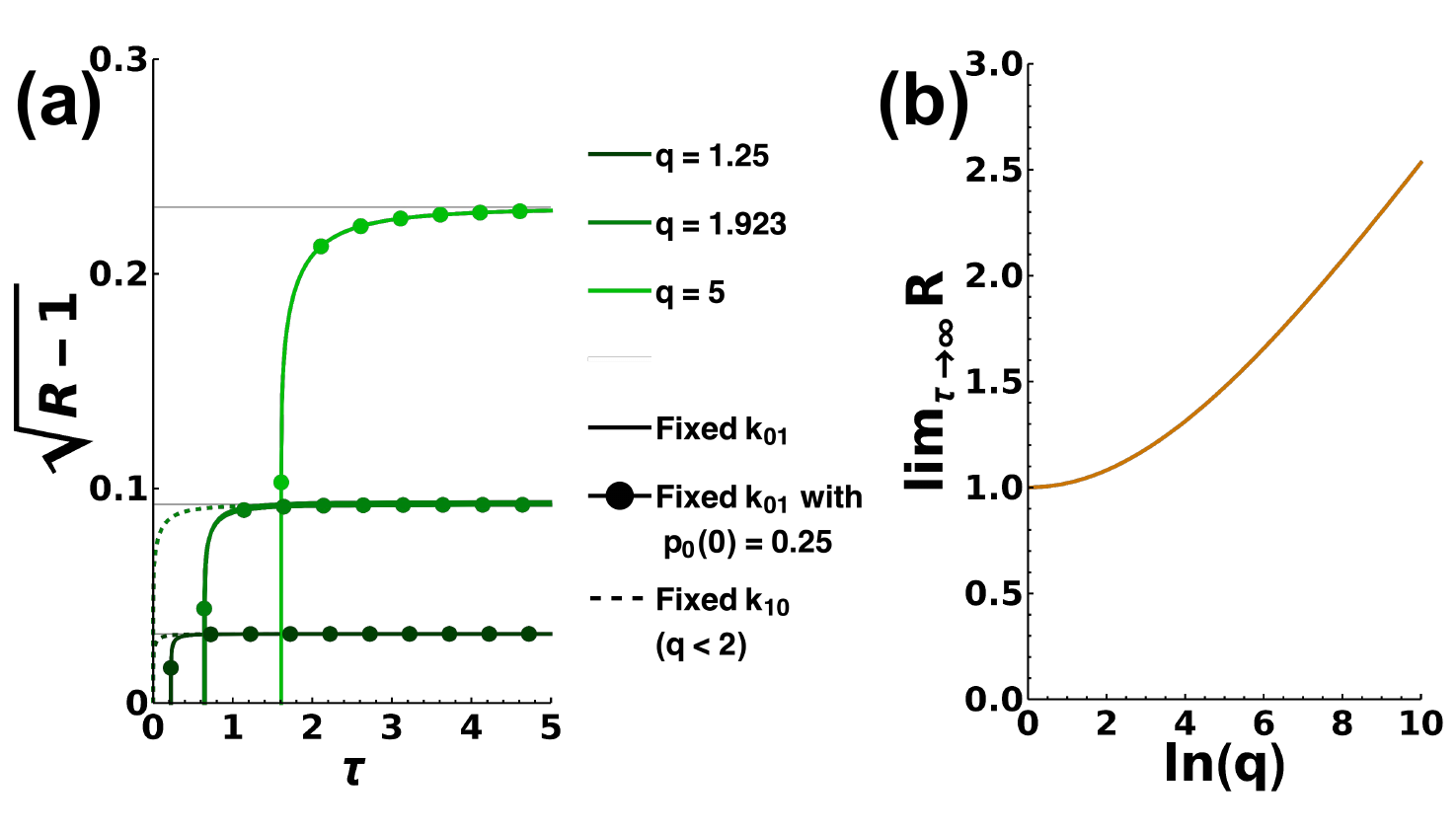}
    \caption{\bl{The change in activity rate over the course of the protocol $q$ determines the ratio between the minimal irreversible work for the two optimization schemes $R$, in the limit $\tau \to \infty$. Panel (a) shows $\sqrt{R-1}$, with $R = W_\text{irr}^\text{min}|_{A_T}/W_\text{irr}^\text{min}|_{\tau}$ (see Eq. (\ref{eq:R})) as a function of $\tau$. Colors correspond to values of $q$, the ratio between the maximal and minimal value of equilibrium activity rate over the course of the protocol (Eq. (\ref{eq:eqA})). For dynamics with $k_{01}=1$, starting symmetrically from $p_0(0)=1/2$, the values $q=1.25$, $q=1.923$ and $q=5$ correspond to an accuracy of $s=0.6$, $s=0.74$ and $s=0.9$, respectively. In the limit $\tau \to \tau_\text{min} = -\log(2(1-s)) = \log(q)$ the two protocols are the same and the ratio becomes 1. For larger values of $\tau$, the ratio increases, and the higher $q$, the larger $\lim_{\tau \to \infty} R$. When the reset is started asymmetrically from $p_0(0) =0.25$ (lines with circles), the results are practically indistinguishable from the symmetric case, when compared for the same value of $q$; the values of $q$ now correspond to $s=0.4$, $s=0.605$ and $s=0.85$, respectively. For an energy-rate relation where $k_{10}$ is kept fixed (dashed lines), $\tau_\text{min}=0$ (see main text) and the maximal value of $q$ is $2$. For $q<2$, $s$ is chosen such that $q$ takes the same values as before ($s=0.625$ and $s=0.95$). For the same value of $q$, $\lim_{\tau \to \infty} R$ is again the same. For a class of dynamics which encompasses both $k_{01}=1$ and $k_{10}=1$, $\lim_{\tau \to \infty} R$ can be expressed as a function of $q$ (Eq. (\ref{eq:limR})). Panel (b) shows this relation as a function of $\ln(q)$. The difference between the two optimization schemes can increase indefinitely with $q$.}}
    \label{fig:scompare}
\end{figure}

\subsection{Using equilibrium activity rate to explain difference between the optimization schemes} 
Optimizing for a given value of $\tau$ is not the same as optimizing for a given value of $A_T$, but what is the reason for this difference? Can we understand the difference in optimal path of the control parameter $\Delta E$, shown in Fig. \ref{fig:worktauA}(c)? Essentially, the two optimization schemes lead to distinct optimal protocols due to the fact that the activity rate \bl{$a(t)$} is both path- and time-dependent. If instead it were always constant, the activity rate for any path would equal \bl{$\langle a \rangle = A_T/\tau$}, and Eq. (\ref{eq:Wmin}) would give the minimal work for both optimization schemes.

\bl{However, \bl{$a(t)$} is not a constant,} for two reasons: when a system is relaxing towards a certain equilibrium state, \bl{$a(t)$} will change due to the changing value of $p_0(t)$ and $p_1(t)= 1-p_0(t)$. Secondly, even for a system which is (approximately) in equilibrium, the activity depends on the equilibrium distribution, so for a protocol with a changing value of $\Delta E(t)$, the activity rate will not be constant, even in the limit of a quasistatic process. In order to understand this second contribution, we approximate \bl{$a(t)$} by its quasistatic value along the path in state space, \bl{$a^\text{eq}(p_0(t))$}, i.e. the activity rate if the energy levels were set to match the current distribution, so that the rates obey $k^\text{eq}_{01}/k^\text{eq}_{10}  = p_0(t)/p_1(t)$: 

\bl{\begin{align}
   a^\text{eq}(p_0(t)) = k^\text{eq}_{10}p_0(t) +k^\text{eq}_{01}p_1(t) = 2 k_{01}p_1(t) = 2 (1-p_0(t)).  
   \label{eq:eqA}
\end{align}}
In this equation, we have exploited that for the protocols considered here the forward rate $k_{01}=1$ is kept constant and only the backward rate $k_{10}(t)$ varies in time. The equilibrium activity rate decreases over the course of the protocol, ranging from \bl{$a^\text{eq}(p_0(0))=1$ to $a^\text{eq}(p_0(\tau))=2(1-s)$.  In general, the relation between $a^\text{eq}$ and $p_0$ depends on the relation between the energy difference and the rates.}

The benefit of singling out the the equilibrium contribution of the activity rate is that its value only depends on time through the probability distribution of the memory, in contrast to the actual activity rate $a(t)$, which depends on the current value of the rate constants (and hence the energy levels) as well. For systems that are not too far from equilibrium, the equilibrium activity rate is a reasonable approximation to the activity rate. In panel (d) of Fig. \ref{fig:worktauA}, we show the equilibrium activity rate for the $\tau$-constrained protocol. Indeed, the behavior of the equilibrium activity rate and the true activity rate (blue solid line) are similar, although the equilibrium activity rate is slightly larger, since the probability distribution lags behind the equilibrium one: $k_{10}(t) < k^\text{eq}_{10}$, and hence \bl{$a(t) < a^\text{eq}(p_0(t))$.}

We are now in the position to explain how the $\tau$-constrained protocol can leverage the variation of the activity rate with time to achieve a lower cost. From Eq. (\ref{eq:eqA}), we see that the dynamics we have imposed lead to an equilibrium activity rate that decreases over the course of the protocol. \bl{In the activity-constrained protocol transitions are penalized. As a result, the optimal protocol tends to move quickly through the regions of state space where the activity rate is high. For the bit reset system considered here, this implies that $\Delta E(t)$ is initially higher and $a(t)$ lower, as compared to the time-constrained protocol. It also means that the irreversible entropy production will initially be higher in the activity-constrained protocol, as compared to the time-constrained one, yet lower later (Fig. \ref{fig:worktauA}(g)). In fact, in the optimal activity-based protocol, $\Delta E(t)$ is such that the entropy production per transition, $\dot{W}_\text{irr}(t) / a(t)$, is constant over the course of the protocol (see Appendix \ref{ss:Kconstant}).} The $\tau$-constrained protocol, on the other hand, uses the high activity to reach a considerable flux, without bringing the system as far out of equilibrium. Indeed, as panel (f) of Fig. \ref{fig:worktauA} shows, the generalized force $\ln(K)$ is initially lower for the $\tau$-constrained protocol. This approach allows the $\tau$-constrained protocol to have a lower cost than the $A_T$-constrained protocol for the same duration $\tau$.

\subsection{Difference between activity- and time-constrained protocol due to variation of activity rate over time}

We now test the hypothesis that the difference between the two protocols is due to change in activity rate over the course of the protocol. In the limit $\tau \to \infty$, the activity rate tends to the equilibrium activity rate, and its relative change over the course of the protocol can be quantified by the ratio

\bl{\begin{align}
    q \equiv \frac{\text{max}_{p_0 \in [p_0(0),p_0(\tau)]} a^\text{eq}(p_0)}{\text{min}_{p_0 \in [p_0(0),p_0(\tau)]} a^\text{eq}(p_0)}. 
    \label{eq:q}
\end{align}}
For our choice of rate-energy relation, the equilibrium activity rate decreases with time, so that \bl{$q = a^\text{eq}(p_0(0))/a^\text{eq}(p_0(\tau))$}. The difference between initial and final equilibrium activity rate increases with $s$, and $q = (2(1-s))^{-1}$. For a system with a different rate-energy relation, like the one where $k_{10}$ is fixed rather than $k_{01}$ (see Fig. \ref{fig:introduction}(c)), the relaxation rate and equilibrium activity rate will increase with time, so that \bl{$q = a^\text{eq}(p_0(\tau))/a^\text{eq}(p_0(0))$}. For ease of comparison later on, $q$ is defined to always be larger than one.

We quantify the difference in minimal irreversible work between both optimization schemes by 

\begin{align}
    R(\tau) \equiv \frac{W_\text{irr}^\text{min}|_{A_T}}{W_\text{irr}^\text{min}|_{\tau}},
    \label{eq:R}
\end{align}
where both the work minimized for $A_T$ and the work minimized for $\tau$ are evaluated for the same value of $\tau$, making $R > 1$, as in the inset of panel (b) in Fig. \ref{fig:worktauA}.  In Fig. \ref{fig:scompare}, we plot $\sqrt{R-1}$ for different values of $s$ as a function of the protocol duration $\tau$ (solid lines), for a low, medium and high vlue of $s$. The values of $s$ shown are $s=0.6$, $s=0.75$ and $s=0.9$ and correspond to a value of $q$ of $q=1.25$, $q=1.923$ and $q=5$, respectively. \bl{We have chosen two values of $q$ smaller than $2$, since we will later compare the results for a fixed forward rate ($k_{01}=1$) to the case of a fixed backward rate ($k_{10}=1$), for which $q_\text{max} =2$.} We plot the square root to increase the visibility of values of $R$ very close to $1$. The ratio starts off at $1$ for $\tau = \tau_\text{min} = -\log(2(1-s))$, and increases to a limiting value in the long-$\tau$ limit. The limiting value increases with $s$, that is, with increasing $q$, as predicted. 

To further probe the relation between the ratio of equilibrium activity rate $q$ and difference between the two optimization schemes, as quantified by $R$, we look at the case of a reset which does not start from a symmetric initial distribution ($p_0(0) = 0.5$), but rather an asymmetric initial distribution $p_0(0)=0.25 $ (line with circles in Fig. \ref{fig:scompare}). We choose $s$ so that the values of $q$ are the same as in the case of the symmetric initial distribution. The graphs are almost indistinguishable: the value of $\tau_\text{min}$, where $R=1$, depends on the ratio of the initial and final distribution which is given by $q$, hence it is the same for both symmetric and asymmetric initial distributions with equal $q$, \bl{see Appendix \ref{ss:Ak01} for a derivation} \cite{Esposito2013}. Furthermore, the value in the limit $\tau \to \infty$ is also the same for the same value of $q$. 

Next, we study the case of a different rate-energy relation: instead of $k_{01}=1$, we set $k_{10}=1$ (dashed lines in Fig. \ref{fig:scompare}). Hence, when $\Delta E$ increases, the rate $k_{01}$ increases, rather than $k_{10}$ decreasing. Since the system can increase its relaxation rate indefinitely (see Fig. \ref{fig:introduction}(b)), there is no minimal time required to perform a certain operation, and $\tau_\text{min}=0$. See Appendix \ref{ss:Ak10} and \ref{ss:tauk10} for the full dynamics and optimal protocols. The equilibrium activity rate \bl{$a^\text{eq}(p_0) = 2 p_0$} now increases with the increasing relaxation rate over the course of a reset, and the ratio $q$ is given by $q = 2 s$ for a reset starting from a symmetric initial distribution. Hence, $q$ can not be greater than $2$. In Fig. \ref{fig:scompare}(a), it can be seen that for the same value of $q$, $R(\tau)$ has the same value in the limit $\tau \to \infty$ for $k_{01}=1$ as for $k_{10}=1$, as long as $q$ is the same. 

In fact, the ratio $R$ can, in the limit $\tau \to \infty$, be quantified solely in terms of $q$, for a general class of dynamics that encompasses both the case $k_{01}=1$ and $k_{10}=1$. For this class of dynamics, \bl{$a^\text{eq}$ is a linear function of $p_0$, so that we can write $a^\text{eq}(p_0) = a_0 + a'p_0$, where $a'$ is the constant derivative of $a^\text{eq}$ with respect to $p_0$. Note that if $k_{01}=1$, then $a_0 = 2$ and $a' = -2$ (see Eq. \ref{eq:eqA}), whereas if $k_{10}=1$, then $a_0=0$ and $a' = 2$.} By methods similar to \cite{sivak12}, we can approximate $W_\text{min}^\text{irr}|_\tau$ in the limit $\tau \to \infty$. The first-order term in $1/\tau$ can be expressed in terms of \bl{$a_0$ and $a'$}, see Appendix \ref{ss:diff}. The activity-constrained minimal work $W_\text{min}^\text{irr}|_{A_T}$ can be expressed in the same limit in terms of \bl{$a_0$, $a'$} and the initial and final distribution, by expressing $A_T$ in terms of $\tau$. Even though these approximations of $W_\text{min}^\text{irr}|_\tau$ and $W_\text{min}^\text{irr}|_{A_T}$ can not be expressed solely in terms of $q$, their ratio $\lim_{\tau \to \infty} R(\tau) \equiv R_\infty$ can, and is given by

\begin{align}
    \lim_{\tau \to \infty} R(\tau)\equiv R_\infty  = \frac{q-1}{4(\sqrt{q}-1)^2}\ln(q).
    \label{eq:limR}
\end{align}
\bl{This result, expressing the ratio of irreversible work for the two optimization schemes as a function of the relative change in activity is reminiscent of results comparing optimal $\tau$-constrained protocols to ``naive" protocols which have not been optimized, in which case the ratio of the work can be written as an expression of the relative change in the friction, related to the inverse of the activity, over the course of the protocol \cite{sivak2016}.} The theoretical prediction is shown in Fig. \ref{fig:scompare}(a) by grey horitzontal gridlines. Indeed, the agreement is excellent: for the same value of $q$, the prediction agrees with the numerically evaluated values in the limit $\tau \to \infty$. In Fig. \ref{fig:scompare}(b), we plot $R_\infty$ as a function of $q$, showing its quadratic behavior in $q-1$ for values of $q \approx 1$ and $\ln(q) \approx (q-1)$, and its logarithmic behavior for $q \to \infty$. Clearly, the difference between the two optimization schemes can be made arbitrarily large by increasing $q$. 

\section{Discussion}  

We have contrasted two optimization schemes for finite-time control of a thermodynamic system described by a Markov model in the context of a bit reset. One optimization scheme minimizes the work to perform the reset operation for a given value of the duration $\tau$. The second one optimizes for a given value of the total activity $A_T$. We show that the $A_T$-constrained scheme is suboptimal for a given $\tau$ and vice versa. Indeed, both schemes lead to different optimal paths. As such, our work is a caveat regarding the work presented in \cite{park22,Vu.2023}, where speed limits are given as a function the product of the average activity rate \bl{$\langle a \rangle$} and the protocol duration $\tau$. We find that for the same duration $\tau$, protocols with a higher value of \bl{$\langle a \rangle$} can have a lower work, despite the fact that they do not saturate the speed limit, nor obey the dynamics of the protocol that is optimal for a constrained total activity \bl{$A_T = \langle a \rangle \tau$}. 

\bl{We study the difference between the two optimization schemes analytically in the long-time limit,} where the activity rate can be approximated by the equilibrium activity rate. Our results show that the difference between the optimization schemes is especially marked if the variation of the activity rate along the path in probability space required for the reset is large. For a general class of dynamics, where the (equilibrium) activity rate is a linear function of the probability distribution, we derive an expression for the difference in work between both protocols in terms of the difference in activity rate. This expression shows that the difference between both optimization schemes can be made arbitrarily large, highlighting the importance of marking the distinction between the two. 

In this paper, we studied the case of a two-state system, but similar considerations most likely hold for more systems with more states. Increasing the number of states drastically increases the complexity of the optimization problem. Firstly, it opens the possibility of non-conservative forces. Worthy of note is the fact that for $\tau$-constrained optimization schemes, there are cases where non-conservative forces are optimal (see \cite{Remlein.2021}), whereas the optimal protocol in the activity-constrained case is based on conservative forces \cite{dechant2022}. Distinguishing between the $\tau$-constrained optimization scheme, where every transition between states obeys a specific relaxation rate-energy relation and hence comes with only a single degree of freedom, and the $A_T$-constrained optimization, where both forward and backward rate of every transition can be chosen freely, under a constraint of total activity over the whole network, is essential to understanding the difference between the two optimal strategies. 

\bl{A second phenomenon that occurs when the number of states is increased,} is that the path in probability space from initial to final distribution will no longer be confined to a single dimension, but will be higher-dimensional and hence no longer unique. Likely, the two optimization schemes will not only differ in terms of the speed at which they travel over the path in probability space, but also in terms of the path that is taken. We predict the difference is especially pronounced if different transitions in the system have markedly different timescales. The duration-constrained optimization scheme will tend to circumvent slower transitions on the network, trading off the fact that more transitions are required, against the fact that transitions will happen at a higher rate. Hence, we deem it unlikely that the minimal cost will be, in general, a function of the Wasserstein distance, like in the $A_T$-constrained case. We note, however, that none of the complexities of multi-state systems, non-conservative forces and different paths through state space are necessary to observe the difference between activity-constrained and duration-constrained protocols.

\section*{Acknowledgements}

We thank Avishek Das, Vahe Galstyan and Claudio Hernandez for a careful reading of the manuscript.
This work is part of the Dutch Research Council (NWO) and was performed
at the research institute AMOLF. This project has received funding from the
European Research Council under the European Union’s Horizon 2020 research
and innovation program (grant agreement No. 885065). TEO is supported by a Royal Society University Research Fellowship.

\newpage
\section{Appendix}
\label{s:appendix}

We present detailed calculations in this Appendix. \bl{First, in section \ref{ss:Kconstant}, we argue why a protocol with a constant of motion given by $K = (k_{01} p_1 )/ (k_{10}p_0)$ is optimal. Next,} we study the case of activity-constrained optimization, as well as time-constrained optimization. For both optimization schemes, we study two options for the relation between relaxation rate and energy difference. In section \ref{ss:Ak01}, we show the optimal protocol for the activity-constrained optimization scheme, given that the forward rate $k_{01}$ is constant. In section \ref{ss:Ak10}, we study the activity-constrained optimization scheme again, but now for the case where the backward rate $k_{10}$ is constant. Also see the Supplemental Information of \cite{park22}, where these equations were introduced and solved. In \ref{ss:tauk01} and \ref{ss:tauk10}, we study the time-constrained optimization scheme for a system with fixed forward rate and backward rate, respectively. Finally, in \ref{ss:diff}, we derive an equation for the ratio between the work given by both optimization schemes in the long-time limit, as a function of the ratio of maximal and minimal activity over the course of the protocol (see Eq. (\ref{eq:limR}) of the main text). 

\subsection{Activity-constrained protocol is optimal for constant $K$}
\label{ss:Kconstant}

\bl{It has been proven that, under a constraint on the total activity, a protocol with a constant of motion given $K = (k_{01}p_1)/(k_{10}p_0)$ is optimal for an $A_T$-constrained protocol. For example, in the Supplemental Information of \cite{park22}, lower bounds are derived on the minimal work for a given total activity, and it is shown that a protocol with a constant of motion given by $K$ can saturate the bound for some value of $K$, making the protocol optimal. Here, we present a different argument, based on a Lagrangian approach, to arrive at the same result. To shorten our notation, we introduce the fluxes $J_{ij} = k_{ij}p_j$, so that the activity rate $a(t) = J_{01}+J_{10}$. The activity-constrained action $\mathcal{S}_{A_T}$ we want to minimize equals}

\bl{\begin{align}
    \mathcal{S}_{A_T} = \int_0^\tau dt \left[ (J_{01}-J_{10}) \ln \left(\frac{J_{01}}{J_{10}} \right)+\lambda (J_{01}+J_{10})\right],
    \label{eq:action}
\end{align}}

\noindent \bl{where the first term equals the entropy production, and the second term in the integrand integrates to the total activity $A_T$, which is penalized by the Lagrange multiplier $\lambda > 0$.}  

\bl{We define the integrand of Eq. (\ref{eq:action}) as the Lagrangian $\mathcal{L}_{A_T}$. The Lagrangian can be rewritten as}
\bl{\begin{align}
    \mathcal{L}_{A_T} = (J_{01}-J_{10})\left(\ln \left(\frac{J_{01}}{J_{10}} \right) + \lambda \frac{\frac{J_{01}}{J_{10}}+1}{\frac{J_{01}}{J_{10}}-1}  \right),
\end{align}}
\noindent\bl{We want to write this Lagrangian in terms of $p_0$ and $\dot{p}_0$. Then, since the Lagrangian is not explicitly time-dependent, we can use the fact that $C = \dot{p_0} \frac{\partial \mathcal{L}_{A_T}}{\partial \dot{p}_0} -\mathcal{L}_{A_T}$ is a constant of motion to find the optimal protocol. We notice that $J_{01}-J_{10} = \dot{p}_0$. In the Lagrangian, this term is multiplied by a function of the ratio of the fluxes $J_{01}/J_{10}$. We now assume that the ratio can be rewritten as a function of the current distribution, parametrized by $p_0$ and its derivative $\dot{p}_0$, i.e. $g(p_0, \dot{p}_0) =  J_{01}/J_{10}$. For the choices of relaxation rates considered in the main text, this assumption holds, see Appendix \ref{ss:tauk01} and \ref{ss:tauk10}. Then, defining $f(x) = (x+1)/(x-1)$, we can write}

\bl{\begin{align}
    C = \dot{p}_0^2\left( \frac{1}{g(p_0,\dot{p}_0)}+\lambda \frac{\partial f\left(g(p_0,\dot{p}_0)\right)}{\partial g(p_0,\dot{p}_0)} \right)\frac{\partial g(p_0,\dot{p}_0)}{\partial \dot{p}_0}.
    \label{eq:argumentmotionSI}
\end{align}}
\bl{Since $1/g(p_0,\dot{p}_0)$ and the derivative of $f$ have opposite sign, there is a simple solution so that $C$ is constant and $0$, namely}

\bl{\begin{align}
\frac{1}{g(p_0,\dot{p}_0)}=-\lambda \frac{\partial f\left(g(p_0,\dot{p}_0)\right)}{\partial g(p_0,\dot{p}_0)}= \frac{2 \lambda}{(g(p_0,\dot{p}_0)-1)^2,}
\end{align}}
\bl{implying that $g(p_0,\dot{p}_0) = J_{01}/J_{10}=K$ is constant.} 

\bl{A similar argument holds for the activity metrics considered in \cite{ito2024}, all of which are means of the fluxes. The activity $a(t) = J_{01}+J_{10}$ is, apart from a factor of $2$, equal to the arithmetic mean of the fluxes, but other choices, like the geometric, harmonic or logarithmic mean, are possible. Any mean $\mu(x,y)$ obeys homogeneity, meaning that $\mu(c x,cy) = c \mu(x,y)$ for $c>0$. Assuming, without loss of generality, that $(J_{01}-J_{10})>0$, $\mu(J_{01},J_{10})$  can be rewritten as $(J_{01}-J_{10})\mu(J_{01}/(J_{01}-J_{10}), J_{10}/(J_{01}-J_{10}))$. That is, the mean can be written as the net flux $\dot{p}_0$ multiplied by a function that only depends on the flux ratio. This fact allows us to write the constant of motion $C$ as in Eq. (\ref{eq:argumentmotionSI}). The functional form of $f$ will depend on the choice of mean $\mu$.}

\bl{Now, from the fact that $K = J_{01}/J_{10}$ is constant, we find that}

\bl{\begin{align}
    A_T = \int_0^\tau dt \dot{p}_0 \frac{K+1}{K-1} =  (p_0(\tau)-p_0(0))\frac{K+1}{K-1.}
\end{align}}
\bl{This expression can be inverted to find}

\bl{\begin{align}
    K = \frac{1+\frac{p_0(\tau)-p_0(0)}{A_T}}{1-\frac{p_0(\tau)-p_0(0)}{A_T}}
\end{align}}
\bl{We then find that} 

\bl{\begin{align}
    W_\text{irr}^\text{min}|_{A_T} &=  \int_0^\tau dt \dot{p}_0 \ln (K) \nonumber \\ &= (p_0(\tau)-p_0(0))\ln \left( \frac{1+\frac{p_0(\tau)-p_0(0)}{A_T}}{1-\frac{p_0(\tau)-p_0(0)}{A_T}}\right),
\end{align}}
\bl{in line with Eq. (\ref{eq:Wmin}). of the main text, where $p_0(0) = 1/2$ and $p_0(\tau)=s$. The ratio $\dot{W}_\text{irr}/a(t) = \ln(K) (J_{01}-J_{10})/(J_{01}+J_{10}) = \ln(K)/f(K)$ is constant for the optimal activity-constrained protocol.}

\subsection{Activity-constrained optimization if $k_{01}$ is constant}
\label{ss:Ak01}

We fix the forward rate, $k_{01}=1$ and study the optimal activity-constrained protocol. Choosing the relation between relaxation rate and energy allows us to explicitly calculate the optimal protocol as a function of time, as well as compare it to the time-constrained optimization scheme. 

For this dynamics, each accuracy comes with a minimal required time. In the limit $\Delta E \to \infty$, the backward rate $k_{10} =0$, so that the master equation (Eq. \ref{eq:mastereq}) reduces to $\dot{p_0} = -\dot{p_1} = p_1$, i.e. $p_1$ decreases exponentially. This is the fastest the system can be operated, and the operation takes 

\begin{align}
    \tau_\text{min} = -\ln \left( \frac{p_1(\tau)}{p_1(0)} \right) = -\ln \left( \frac{1-p_0(\tau)}{1-p_0(0)} \right). 
\end{align}
This result agrees with the case discussed in the main text of a symmetric initial distribution, where $\tau_\text{min} = -\ln(2(1-s))$.

We now calculate the optimal protocol and the minimal required work for a reset which starts at $p_0(0)$ and ends at $p_0(\tau)$. In the symmetrical case discussed in the main text, $p_0(0) =1/2$ and $p_0(\tau) = s$, where $s$ is the accuracy of the reset. We combine the master equation

\begin{align}
    \dot{p}_0 = k_{01}p_1 - k_{10}p_0
    \label{eq:mastereqplainSI}
\end{align}
and the constant of motion \cite{park22}

\begin{align}
    K = \frac{k_{01}p_1}{k_{10}p_0}
    \label{eq:Kdefsi}
\end{align}
with the fixed forward rate $k_{01}=1$, to find

\begin{align}
    \dot{p_1} = - (K-1) k_{10}p_0 = -\frac{K-1}{K}p_1,
\end{align}
so that 

\begin{align}
    p_1(t) = p_1(0) e^{-\frac{K-1}{K}t}
\label{eq:dynamicsp1}
\end{align}
and

\begin{align}
    p_0(t) = 1- (1-p_0(0)))e^{-\frac{K-1}{K}t}.
    \label{eq:p0}
\end{align}
At $t=\tau$, Eq. (\ref{eq:p0}) can be rewritten as

\begin{align}
    \tau = \frac{K}{K-1}\log \left(\frac{1-p_0(0)}{1-p_0(\tau)} \right),
\end{align}
or, vice versa

\begin{align}
    K = \frac{\tau}{\tau-\log \frac{1-p_0(0)}{1-p_0(\tau)}}.
    \label{eq:Ksi1}
\end{align}

We can express the activity rate as 

\bl{\begin{align}
    a(t) &= k_{01}p_1+k_{10}p_0 = (1+\frac{1}{K})k_{01}p_1 \nonumber \\
    &= (1-p_0(0)) \frac{K+1}{K} e^{-\frac{K-1}{K}t}. 
\end{align}}
Integrating the activity rate, we find that

\bl{\begin{align}
    A_T &= \int_0^\tau a(t) dt = (1-p_0(0))\frac{K+1}{K-1}(1-e^{-\frac{K-1}{K}\tau}) \nonumber\\
    &= 2 (p_0(\tau) - p_0(0))\frac{\tau- \frac{1}{2}\log \left( \frac{1-p_0(0)}{1-p_0(\tau)} \right)}{\log \left( \frac{1-p_0(0)}{1-p_0(\tau)} \right)}
    \label{eq:AtSI1}.
\end{align}}
Inversely, 

\begin{align}
    \tau =  \log \left(\frac{1-p_0(0)}{1-p_0(\tau)}\right) \left(\frac{1}{2}+\frac{A_T}{2(p_0(\tau)-p_0(0))}\right).
\end{align}
The minimal irreversible work equals

\begin{align}
    W^\text{min}_\text{irr}|_{A_T} &= \int_0^\tau dt \dot{p}_0 \ln \left( \frac{k_{01}p_1}{k_{10}p_0} \right) = \ln(K) \int_0^\tau dt 
\dot{p}_0 \nonumber \\
    & = (p_0(\tau)-p_0(0))\ln(K).
\end{align}
Using the relation for $K$, and $\tau$, the minimal irreversible work can be expressed in terms of $A_T$, $p_0(0)$ and $p_1(\tau)$ as

\begin{align}
    W^\text{min}_\text{irr}|_{A_T} = (p_0(\tau)-p_0(0))\ln \left( \frac{1 + \frac{p_0(\tau)-p_0(0)}{A_T}}{1- \frac{p_0(\tau)-p_0(0)}{A_T}}\right),
    \label{eq:WmingeneralSI}
\end{align}
congruent with Eq. \ref{eq:Wmin} of the main text, where $p_0(0) = \frac{1}{2}$ and $p_0(\tau) = s$. Lastly, the optimal path of the control parameter $\Delta E(t)$ can be found as follows

\begin{align}
    \Delta E(t) &= \log \left(\frac{k_{01}}{k_{10}} \right) = \log (K) + \log \left( \frac{p_0}{p_1} \right)\nonumber \\
    &=\log(K) + \log \left(\frac{1-(1-p_0(0)))e^{-\frac{K-1}{K}t}}{(1-p_0(0))e^{-\frac{K-1}{K}t}} \right).
\end{align}\\

\subsection{Activity-constrained optimization if $k_{10}$ is constant}
\label{ss:Ak10}

In the case that $k_{10} =1$, we can straightforwardly redo the same calculation as shown in the previous paragraph for the case $k_{01}=1$. These calculations lead to

\begin{align}
    p_0(t) = p_0(0)e^{(K-1)t},
\end{align}

\begin{align}
    \tau = \frac{\log\left(\frac{p_0(\tau)}{p_0(0)}\right)}{K-1},
\end{align}

\begin{align}
    K = 1 + \frac{\log \left( \frac{p_0(\tau)}{p_0(0)}\right)}{\tau},
    \label{eq:Ksi2}
\end{align}

\bl{\begin{align}
    a(t) = p_0(0)(K+1)e^{(K-1)t},
\end{align}}

\begin{align}
    A_T &= p_0(0) \left( \frac{K+1}{K-1} \right)(e^{(K-1)\tau}-1) \nonumber \\
    &= 2 \left(p_0(\tau)-p_0(0)\right) \frac{\tau + \frac{1}{2}\log \left(\frac{p_0(\tau)}{p_0(0)}\right)}{\log \left( \frac{p_0(\tau))}{p_0(0)}\right)}.
    \label{eq:Asi2}
\end{align}
We can invert the linear relation for $A_T$ to find

\begin{align}
    \tau = \left(\frac{A_T}{2 \left( p_0(\tau)-p_0(0)\right)}-\frac{1}{2}\right)\log \left(\frac{p_0(\tau)}{p_0(0)} \right).
\end{align}
Furthermore, 

\begin{align}
    W^\text{min}_\text{irr}|_{A_T} = (p_0(\tau)-p_0(0))\ln(K),
\end{align}
and

\begin{align}
    \Delta E (t) = \ln(K) + \ln \left( \frac{p_0(0)e^{(K-1)t}}{1-p_0(0)e^{(K-1)t}}\right).
\end{align}\\

Notably, the results for the case $k_{10}=1$ are the same as the ones for the case $k_{01}=1$, under the substitution $K \to 1/K$  and $1-p_0 =p_1 \to p_0$. This symmetry can be understood as a relabeling of the states $0$ and $1$, which turns $K$ (Eq. (\ref{eq:Kdefsi})) into $K' = 1/K$, which is also a constant of motion. Furthermore, the symmetry operation changes which of the rates is kept constant and changes $p_0$ into $p_1 = 1-p_0$ . Hence, using the results for $k_{01}=1$ from the previous section for a reset towards state $1$ (decreasing $p_0$) will be the same as the results from this section for a reset towards state $0$ (increasing $p_0$).

\subsection{Deriving $W^\text{min}_\text{irr}|_\tau$ for fixed $k_{01}$}
\label{ss:tauk01}

As first presented in \cite{Seifert_2014}, we can derive the optimal protocol to change the distribution of a bit from the initial distribution parameterized by $p_0(0)$ to a final distribution $p_0(\tau)$ within a time $\tau$, and the corresponding minimal work $W^\text{min}_\text{irr}|_\tau$, for a system which has one of its rates fixed. For a system with $k_{01} = 1$, the dynamics are

\begin{align}
    \dot{p_0} = p_1-k_{10}p_0 = 1-(1+k_{10})p_0. 
\end{align}
We can use this relation to express the irreversible work of Eq. (\ref{eq:Wirr}) in terms of $p_0$, $\dot{p}_0$ and constants, and define a corresponding Lagrangian density so that minimizing this Lagrangian minimizes the irreversible work. The Lagrangian is

\begin{align}
    \mathcal{L}(p_0, \dot{p_0}) = \dot{p}_0 \ln \left( \frac{1-p_0}{1-p_0 - \dot{p}_0}\right).
    \label{eq:lagrangeSI}
\end{align}
Since the Lagrangian has no explicit time dependence, it has a conserved quantity $\kappa = \dot{p_0} \frac{\partial \mathcal{L}}{\partial \dot{p}_0} - \mathcal{L}$, which yields: 

\begin{align}
    \kappa = \frac{\dot{p}_0^2}{1-p_0 - \dot{p_0}},
    \label{eq:kappaSI}
\end{align}
as in the main text. 

The formula for $\kappa$ in Eq. (\ref{eq:kappaSI}) is a quadratic equation in $\dot{p_0}$ which has the following positive solution:

\begin{align}
    \dot{p_0} &= \frac{-\kappa + \sqrt{\kappa^2 + 4 \kappa (1-p_0)}}{2} \nonumber \\
    &= \frac{\kappa}{2}(\sqrt{1+4(1-p_0)/\kappa}-1).
    \label{eq:p0dot}
\end{align}
We use the positive solution since we study the case of reset, where $p_0$ is increasing. We can use this expression for $\dot{p_0}$ to express the irreversible work in terms of just $p_0$, and subsequently calculate $W$ using Eq. (\ref{eq:Wirr}) by changing variables from $t$ to $p_0(t)$ when integrating Eq. (\ref{eq:lagrangeSI}):

\begingroup\makeatletter\def\f@size{9}\check@mathfonts
\def\maketag@@@#1{\hbox{\m@th\large\normalfont#1}}%
\begin{align}
    W^\text{min}_\text{irr}|_\tau &= \int_{0}^{\tau}dt \mathcal{L} =  \int_{p_0(0)}^{p_0(\tau)}dp_0 \ln\left( \frac{1-p_0}{1-\dot{p_0}-p_0}\right) \nonumber \\
    &=\int_{p_0(0)}^{p_0(\tau)}dp_0 \ln \left(\frac{1- p_0}{1 - p_0- \frac{\kappa}{2}(\sqrt{1+4(1-p_0)/\kappa}-1)} \right),
    \label{eq:Wmintau1}
\end{align}\endgroup
which can be solved to yield

\begin{align}
    W^\text{min}_\text{irr}|_\tau = \left[ -\frac{\kappa}{2}\sqrt{Y} - (1-p_0)\ln \left( \frac{\sqrt{Y}+1}{\sqrt{Y}-1} \right) \right]_{p_0(0)}^{p_0(\tau)},
\end{align}
with $Y = 1+4 (1-p_0)/\kappa$ defined for brevity.

To calculate $W^\text{min}_\text{irr}|_\tau$ as a function of $p_0(0)$, $p_0(\tau)$ and $\tau$, we need to determine the appropriate value of $\kappa$. In order to do so, we solve the differential equation for $p_0$. We return to Eq. (\ref{eq:p0dot}), which by separation of variables and integrating can be rewritten as

\begin{align}
    \int_{p_0(0)}^{p_0(t)} dp_0 \frac{2}{\sqrt{\kappa^2+4 \kappa(1-p_0)}-\kappa} =t.
    \label{eq:intunsolved}
\end{align}

\noindent The integral can be solved to yield

\begin{align}
   t = \Big[ &-  \sqrt{1+4(1-p_0(t))/\kappa}  \nonumber\\ 
   &- \ln \left({\sqrt{1+4(1-p_0(t))/\kappa}-1} \right)\Big]_{p_0(0)}^{p_0(t)}.
\label{eq:solvetauSI}
\end{align}
Filling in $p_0(0)$,  $p_0(\tau)$ and $t = \tau$ gives us the equation $\kappa$ must obey for the operation to have the desired properties. Since Eq. (\ref{eq:solvetauSI}) is a transcendental equation, $W^\text{min}_\text{irr}|_\tau$ can not simply be written in terms of $p_0(0)$, $p_0(\tau)$ and $\tau$. However, the system of equations can be solved numerically to find $W_\text{irr}^\text{min}|_{\tau}$.\\

\subsection{Deriving $W^\text{min}_\text{irr}|_\tau$ for fixed $k_{10}$}
\label{ss:tauk10}

In the case that $k_{10} = 1$, the dynamics are

\begin{align}
    \dot{p}_0 =k_{01} p_1 - p_0.
\end{align}
Expressed in terms of $p_1$, we have

\begin{align}
    \dot{p}_1 = 1- (1+k_{01})p_1.
\end{align}
We can rewrite $W_\text{irr}$ in Eq. (\ref{eq:Wirr}) using the relation above, as

\begin{align}
    W_\text{irr} &= \int_0^\tau dt \dot{p_1} \ln \left( \frac{k_{10}p_0}{k_{01}p_1}\right) \nonumber\\
    &= \int_0^\tau dt \dot{p_1} \ln \left( \frac{1-p_1}{1-p_1 -\dot{p_1}}\right).
\end{align}
Comparing this expression to Eq. (\ref{eq:lagrangeSI}), we see that when $k_{10}=1$, $p_1$ obeys the same optimal dynamics as $p_0$ in the case that $k_{01}=1$, much like the symmetry in the case of optimizing for a constrained activity. However, we can not simply replace $p_0$ with $p_1$ in the equations for $W_\text{irr}^\text{min}|_\tau$ derived for the case $k_{10}=1$, since we made use of the fact that $p_0$ is increasing. Instead, $p_1$ decreases, so

\begin{align}
    \dot{p_1} = -\frac{\kappa'}{2}\left(\sqrt{1+4(1-p_1)/\kappa'}+1\right),
\end{align}
where $\kappa'$ is defined as $\kappa$ in Eq. (\ref{eq:kappaSI}) but with $p_0$ replaced by $p_1$. Rewriting leads to 

\begin{align}
   W^\text{min}_\text{irr}|_\tau = \left[ \frac{\kappa'}{2} \sqrt{Y'} +(1-p_1)\ln \left( \frac{\sqrt{Y'}+1}{\sqrt{Y'}-1}\right)\right]_{p_1(0)}^{p_1(\tau)},
\end{align}
where $Y'= 1+4 (1-p_1)/\kappa'$. We can determine $\kappa '$ in the above expression by solving

\begin{align}
    t = \left[ \sqrt{Y'} - \ln \left( \sqrt{Y'} +1\right) \right]_{p_1(0)}^{p_1(t)},
\end{align}
in the case $t=\tau$, using in the boundary conditions $p_1(0)$ and $p_1(\tau)$, see also Eq. \ref{eq:Wmintau1}-\ref{eq:solvetauSI}.  This result is a transcendental equation for $\kappa'$, which can be solved numerically. The resulting $\kappa$ can be used to find $W_\text{irr}^\text{min}|_\tau$.\\

\subsection{Approximating the difference between both optimization schemes in the long-$\tau$ limit for general dynamics}
\label{ss:diff}

We hypothesize that the difference between the activity-constrained and time-constrained optimization scheme is due to the fact that the activity rate is not constant in time, even in the case of long duration $\tau$, when the operation is approximately quasistatic. In this quasistatic limit, the time-dependence of the activity rate is due to the fact that the activity rate depends on the equilibrium distribution $p^\text{eq}_0(t)$, which changes with $\Delta E(t)$. We quantify the difference between both optimization schemes by the ratio of the minimal work for both optimization schemes, and study the quasistatic limit

\begin{align}
    \lim_{\tau \to \infty} R(\tau) = R_\infty = \lim_{\tau \to \infty} \frac{W^\text{min}_\text{irr}|_{A_T}}{W^\text{min}_\text{irr}|_{\tau}}. 
\end{align} 
Note that, even though $W^\text{min}_\text{irr}|_{A_T}$ and $W^\text{min}_\text{irr}|_{\tau}$ are both $0$ in the quasistatic limit ($\tau \to \infty$ and $A_T \to \infty$), their ratio is not, as panel (a) of Figure \ref{fig:scompare} of the main text shows. In the quasistatic limit, the activity rate can be approximated by the equilibrium activity rate, as defined in Eq. (\ref{eq:eqA}) of the main text, where the current distribution is assumed to be the equilibrium one, with rates complying with the latter. The variation in equilibrium activity rate can then be quantified by the ratio between maximal and minimal value of the equilibrium activity rate

\bl{\begin{align}
    q = \frac{\text{max}_{p_0 \in [p_0(0), p_0(\tau)]}(a^\text{eq}(p_0))}{\text{min}_{p_0 \in [p_0(0), p_0(\tau)]}(a^\text{eq}(p_0))},
\end{align}}
Note that this quantity is defined to be greater than or equal to one. For the monotonic relations between $p_0$ and \bl{$a^\text{eq}$} considered here, $q$ gives the ratio of the equilibrium activity at the start and endpoint of the protocol: the value at the start is divided by the value at the end (or vice versa) when \bl{$a^\text{eq}$} decreases (increases) with $p_0$. 

We want to find an expression for $R_\infty$ as a function of $q$. We could use the results derived in the sections above on the activity-constrained and time-constrained optimization to express $R_\infty$. However, those results will then be specific to the choices between energy and relaxation rate, where either $k_{01}$ or $k_{10}$ is fixed. Instead, we present a result for a more general class of relations between relaxation rate and energy difference, which encompasses the case $k_{01}=1$ and $k_{10}=1$. In this class, the equilibrium activity rate is a linear function of the probability distribution: \bl{$a^\text{eq}(p_0(t)) = a_0 + a' p_0(t)$, where $a_0$ and $a'$} are both constants. Note that the case $k_{01}=1$, 

\bl{\begin{align}
    a^\text{eq}(p_0) &= k^\text{eq}_{01}(1-p_0)+k_{10}^\text{eq}p_0 = (1-p_0) + \frac{1-p_0}{p_0}p_0 \nonumber \\
    &= 2(1-p_0),
\end{align}}
so \bl{$a_0 =2 $ and ${a'}=-2$. Similarly, the case $k_{10}=1$ corresponds to $a_0 =0 $ and $a'=2$.} We will show that, regardless of the values of \bl{$a_0$ and $a'$}, $R_\infty$ is given by the same, increasing function of $q$. 

In order to find $R_\infty$, we first approximate the activity-constrained minimal work $W^\text{min}_\text{irr}|_{A_T}$ in the long $A_T$-limit. Subsequently, we express $A_T$ in terms of $\tau$, to make the comparison on the basis of equal $\tau$. Lastly, we approximate $W^\text{min}_\text{irr}|_{\tau}$ in the long-$\tau$ limit. We then have all the ingredients to calculate $R_\infty$.

We approximate the activity-constrained minimal work $W^\text{min}_\text{irr}|_{A_T}$ (see Eq. (\ref{eq:WmingeneralSI})) in the large $A_T$-limit by Taylor expanding the logarithm as

\begin{align}
    W^\text{min}_\text{irr}|_{A_T} \approx \frac{2(p_0(\tau) - p_0(0))^2}{A_T}.
\label{eq:WminATlongATSI}
\end{align}
In order to compare this expression to $W^\text{min}_\text{irr}|_{\tau}$, we need to express $A_T$ in Eq. (\ref{eq:WminATlongATSI}) in terms of $\tau$. In the slow-protocol limit, $p_0 \approx p_0^\text{eq}$, and we can approximate the relevant quantities to calculate $A_T$ ($\dot{p_0}$, $K$ and \bl{$a(t)$}) to first order in $p_0 - p_0^\text{eq}$. The constant of motion $K$ (\ref{eq:Kdefsi}) is approximated as

\begin{align}
    K = \frac{k_{01}p_1}{k_{10}p_0} = \frac{p_0^\text{eq}(1-p_0)}{(1-p_0^\text{eq})p_0} \approx 1 - \frac{p_0 - p_0^\text{eq}}{(1-p_0^\text{eq})p_0^\text{eq}}.
\end{align}
The equilibrium activity rate can be expressed as

\bl{\begin{align}
    a^\text{eq} &= a_0 + a' p^\text{eq}_0 = k_{01} (1-p^\text{eq}_0)+k_{10} p^\text{eq}_0 \nonumber \\
    &= 2 (k_{10} + k_{01})p_0^\text{eq} (1-p_0^\text{eq}), 
    \label{eq:ASI}
\end{align}}
which we use to approximate the master equation as

\bl{\begin{align}
    \dot{p}_0 &= k_{01}p_1 - k_{10}p_0 =  -(k_{10} + k_{01})(p_0 - p_0^\text{eq}) \nonumber\\
 &=  \bl- \frac{a_0 + a'p_0^\text{eq}}{2 p_0^\text{eq}(1-p_0^\text{eq})}(p_0 - p_0^\text{eq}) \nonumber \\
&\approx \frac{1}{2}(K-1)(a_0 + a'p_0^\text{eq}) \approx \frac{1}{2}(K-1)(a_0 + a'p_0).
\label{eq:mastereqSI}
\end{align}}
In the last line, we have used the fact that $K-1$ is small to approximate $p_0^\text{eq}$ by $p_0$. The differential equation Eq. \ref{eq:mastereqSI} is solved by  

\bl{\begin{align}
    p_0(t) = \left(a_0 + a'p_0(0) \right) e^{\frac{1}{2}(K-1)a't} - \frac{a_0}{a'},
\end{align}}
which, substituting in $t = \tau$, leads to the following expression for $(K-1)$:

\bl{\begin{align}
    K-1 = \frac{2}{a'\tau} \ln \left(\frac{a_0 + a' p_0(\tau)}{a_0+a'p_0(0)} \right).
\end{align}}
\bl{For the appropriate values of $a_0$ and $a'$, i.e. $a_0=2$ and $a'=-2$ when $k_{01}=1$, and $a_0 = 0$ and $a'=2$ when $k_{10}=1$}, this expression indeed agrees with the value of $K$ found earlier for either fixed $k_{01}$ (Eq. (\ref{eq:Ksi1})) or $k_{10}$ (Eq. (\ref{eq:Ksi2})) in the large $\tau$-limit.

Now, $A_T$ can be calculated by integrating \bl{$a(t)$}. We approximate the activity rate to zeroth order in $p_0 - p_0^\text{eq}$, essentially using \bl{$a^\text{eq}(t)$}, which can be expressed in terms of $\dot{p_0}$ using Eq. (\ref{eq:mastereqSI}) as

\bl{\begin{align}
    a(t) &\approx a^\text{eq}(t) = a_0 + a' p_0= \frac{2 \dot{p_0}}{K-1},
\end{align}}
so that

\begin{align}
    A_T &= \int_0^\tau dt \bl{a(t)} \approx \frac{2\left(p_0(\tau) - p_0(0)\right)}{K-1} \nonumber\\
    &= \frac{\bl{a'} \tau \left(p_0(\tau) - p_0(0)\right)}{\ln \left( \frac{\bl{a_0} + \bl{a'}p_0(\tau)}{\bl{a_0} + \bl{a'}p_0(0)} \right)}.
    \label{eq:AT3SI}
\end{align}
Eq. (\ref{eq:AT3SI}) agrees with earlier results (see Eq. (\ref{eq:AtSI1}) and (\ref{eq:Asi2})) for the appropriate values of \bl{$a_0$ and $a'$} in the long-$\tau$ limit. Given this expression for $A_T$, $W_\text{irr}^\text{min}|_{A_T}$ in Eq. (\ref{eq:WminATlongATSI}) can be expressed in terms of $\tau$ as

\bl{\begin{align}
     W^\text{min}_\text{irr}|_{A_T} = \frac{2(p_0(\tau) - p_0(0))}{a'\tau}\ln \left( \frac{a_0 + a' p_0(\tau)}{a_0+ a'p_0(0)} \right).
     \label{eq:WminATfinalSI}
\end{align}}

Next, we will express $W_\text{irr}^\text{min}|_{\tau}$ in limit $\tau \to \infty$ to first order in $1/\tau$. In order to do so, we note that we can rewrite $W_\text{irr}$ (Eq. (\ref{eq:Wirr})) by using $k_{01}/k_{10} = p^\text{eq}_0/p^\text{eq}_1$ and $p_1 = 1-p_0$ as

\begin{align}
    W_\text{irr} = \int_0^\tau dt \dot{p_0} \ln \left(\frac{p_0^\text{eq}(1-p_0)}{(1-p_0^\text{eq})p_0} \right).
    \label{eq:WirrrewrittenSI}
\end{align}
The integrand can be interpreted as a Lagrangian density. In order to minimize Eq. (\ref{eq:WirrrewrittenSI}) using the Euler-Lagrange formalism, we want to eliminate $p_0^\text{eq}$ and express the integrand in terms of $p_0$ and $\dot{p_0}$. We can use the master equation Eq. (\ref{eq:mastereqplainSI}) to express $p_0^\text{eq}$ in terms of the other two variables, by rewriting it using $p_0^\text{eq} = k_{01}/(k_{01}+k_{10})$ as

\begin{align}
    \dot{p}_0 = -k_T(p_0^\text{eq})(p_0 - p_0^\text{eq})
    \label{eq:mastereqSI2}
\end{align}
where $k_T = k_{01}+k_{10}$. How $k_T$ depends on $p_0^\text{eq}$ depends on the chosen relation between energy and rates. For example, when $k_{01} = 1$, $k_T(p_0^\text{eq}) = 1 + (1-p_0^\text{eq})/p_0^\text{eq}$, and when  $k_{10} = 1$, $k_T(p_0^\text{eq}) = 1 + p_0^\text{eq}/(1-p_0^\text{eq})$. As shown in the sections where the $\tau$-constrained work is derived for these specific cases, they have the property that Eq. (\ref{eq:mastereqSI2}) can be easily inverted to find an expression for $p_0^\text{eq}$. However, inverting Eq. (\ref{eq:mastereqSI2}) generally leads to expressions that are unwieldy. Instead, since we are interested in the limit $\tau \to \infty$ for which $p_0 - p_0^\text{eq}$ is small, we approximate the dynamics (Eq. (\ref{eq:mastereqSI2})) to first order in $p_0 - p_0^\text{eq}$: 

\begin{align}
    \dot{p_0} \approx -k_T(p_0)(p_0 - p_0^\text{eq}).
\end{align}
This expression can be straightforwardly solved for $p_0^\text{eq} = p_0 + \dot{p_0}/k_T(p_0)$, so that 

\bl{\begin{align}
    W_\text{irr} &\approx \int_0^\tau dt \dot{p_0} \ln \left(\frac{(1+p_0 + \frac{\dot{p_0}}{k_T(p_0)})(1-p_0)}{(1-p_0 - \frac{\dot{p_0}}{k_T(p_0)})p_0} \right) \nonumber\\
    &\approx \int_0^\tau dt \frac{\dot{p_0}^2}{k_T(p_0) p_0 (1-p_0)} \nonumber\\
&\approx   \int_0^\tau dt \frac{\dot{2 p_0}^2}{a_0 + a'p_0},
    \label{eq:WirrrewrittenSI2}
\end{align}}
where in the last step, Eq. (\ref{eq:ASI}) is used to express the rates in terms of the activity. 

The integrand of Eq. (\ref{eq:WirrrewrittenSI2}) can be interpreted as a Langrangian density. Since it has no explicit time-dependence, the constant of motion $\kappa$ for the optimal trajectory is 

\bl{\begin{align}
    \kappa = \dot{p_0} \frac{\partial \mathcal{L}}{\partial \dot{p_0(0)}} - \mathcal{L} = \frac{2 \dot{p_0}^2}{a_0 + a' p_0 }.
\label{eq:kappaSI2}
\end{align}}
Hence, $W_\text{irr}^\text{min}|_\tau = \kappa \tau$. We determine $\kappa$ by integrating the square root of the left and right hand side of Eq. (\ref{eq:kappaSI2}), to find

\bl{\begin{align}
    \sqrt{\kappa}\tau &= \int_0^\tau dt \frac{\sqrt{2}\dot{p_0}}{\sqrt{a_0 + a'p_0}} = \left[ \frac{2 \sqrt{2} \sqrt{a_0 + a'p_0}}{a'}\right]_{p_0(0)}^{p_0(\tau)} \nonumber \\
    &= \frac{2\sqrt{2}}{a'}\left(\sqrt{a_0 + a'p_0(\tau)} - \sqrt{a_0 + a'p_0(0)}\right),
\end{align}}
so that

\bl{\begin{align}
    W_\text{irr}^\text{min}|_\tau = \frac{8}{a'^2\tau}\left(\sqrt{a_0 + a'p_0(\tau)} - \sqrt{a_0 + a'p_0(0)} \right)^2.
    \label{eq:WmintaufinalSI} 
\end{align}}

The ratio $R_\infty$ is given by the ratio of Eq. (\ref{eq:WminATfinalSI}) and Eq. (\ref{eq:WmintaufinalSI}), and equals

\bl{\begin{align}
    R_\infty &= \frac{a'(p_0(\tau) - p_0(0))}{4\left(\sqrt{a_0 + a'p_0(\tau)} - \sqrt{a_0 + a'p_0(0)} \right)^2} \ln \left( \frac{p_0(\tau)+ \frac{a_0}{a'}}{p_0(0)+\frac{a_0}{a'}} \right) \nonumber\\
    &=\frac{a^\text{eq}(\tau) - a^\text{eq}(0) }{4(\sqrt{a^\text{eq}(\tau)}-\sqrt{a^\text{eq}(0)})^2}\ln \left( \frac{a^\text{eq}(\tau)}{a^\text{eq}(0)} \right),
\end{align}}
where we used \bl{$a_0 + a'p_0 = a^\text{eq}$. Using the definition of $q= \text{max}(a^\text{eq}(t))/\text{min}(a^\text{eq}(t))$}, $R_\infty$ can be written as

\begin{align}
    R_\infty = \frac{q-1}{4(\sqrt{q}-1)^2}\ln(q). 
\end{align}


\begin{thebibliography}{39}%
\makeatletter
\providecommand \@ifxundefined [1]{%
 \@ifx{#1\undefined}
}%
\providecommand \@ifnum [1]{%
 \ifnum #1\expandafter \@firstoftwo
 \else \expandafter \@secondoftwo
 \fi
}%
\providecommand \@ifx [1]{%
 \ifx #1\expandafter \@firstoftwo
 \else \expandafter \@secondoftwo
 \fi
}%
\providecommand \natexlab [1]{#1}%
\providecommand \enquote  [1]{``#1''}%
\providecommand \bibnamefont  [1]{#1}%
\providecommand \bibfnamefont [1]{#1}%
\providecommand \citenamefont [1]{#1}%
\providecommand \href@noop [0]{\@secondoftwo}%
\providecommand \href [0]{\begingroup \@sanitize@url \@href}%
\providecommand \@href[1]{\@@startlink{#1}\@@href}%
\providecommand \@@href[1]{\endgroup#1\@@endlink}%
\providecommand \@sanitize@url [0]{\catcode `\\12\catcode `\$12\catcode `\&12\catcode `\#12\catcode `\^12\catcode `\_12\catcode `\%12\relax}%
\providecommand \@@startlink[1]{}%
\providecommand \@@endlink[0]{}%
\providecommand \url  [0]{\begingroup\@sanitize@url \@url }%
\providecommand \@url [1]{\endgroup\@href {#1}{\urlprefix }}%
\providecommand \urlprefix  [0]{URL }%
\providecommand \Eprint [0]{\href }%
\providecommand \doibase [0]{http://dx.doi.org/}%
\providecommand \selectlanguage [0]{\@gobble}%
\providecommand \bibinfo  [0]{\@secondoftwo}%
\providecommand \bibfield  [0]{\@secondoftwo}%
\providecommand \translation [1]{[#1]}%
\providecommand \BibitemOpen [0]{}%
\providecommand \bibitemStop [0]{}%
\providecommand \bibitemNoStop [0]{.\EOS\space}%
\providecommand \EOS [0]{\spacefactor3000\relax}%
\providecommand \BibitemShut  [1]{\csname bibitem#1\endcsname}%
\let\auto@bib@innerbib\@empty
%</preamble>
\bibitem [{\citenamefont {Salamon}\ and\ \citenamefont {Berry}(1983)}]{Berry83}%
  \BibitemOpen
  \bibfield  {author} {\bibinfo {author} {\bibfnamefont {Peter}\ \bibnamefont {Salamon}}\ and\ \bibinfo {author} {\bibfnamefont {R.~Stephen}\ \bibnamefont {Berry}},\ }\bibfield  {title} {\enquote {\bibinfo {title} {Thermodynamic length and dissipated availability},}\ }\href {\doibase 10.1103/PhysRevLett.51.1127} {\bibfield  {journal} {\bibinfo  {journal} {Phys. Rev. Lett.}\ }\textbf {\bibinfo {volume} {51}},\ \bibinfo {pages} {1127--1130} (\bibinfo {year} {1983})}\BibitemShut {NoStop}%
\bibitem [{\citenamefont {Crooks}(2007)}]{crooks07}%
  \BibitemOpen
  \bibfield  {author} {\bibinfo {author} {\bibfnamefont {Gavin~E.}\ \bibnamefont {Crooks}},\ }\bibfield  {title} {\enquote {\bibinfo {title} {Measuring thermodynamic length},}\ }\href {\doibase 10.1103/PhysRevLett.99.100602} {\bibfield  {journal} {\bibinfo  {journal} {Phys. Rev. Lett.}\ }\textbf {\bibinfo {volume} {99}},\ \bibinfo {pages} {100602} (\bibinfo {year} {2007})}\BibitemShut {NoStop}%
\bibitem [{\citenamefont {Sivak}\ and\ \citenamefont {Crooks}(2012)}]{sivak12}%
  \BibitemOpen
  \bibfield  {author} {\bibinfo {author} {\bibfnamefont {David~A.}\ \bibnamefont {Sivak}}\ and\ \bibinfo {author} {\bibfnamefont {Gavin~E.}\ \bibnamefont {Crooks}},\ }\bibfield  {title} {\enquote {\bibinfo {title} {Thermodynamic metrics and optimal paths},}\ }\href {\doibase 10.1103/PhysRevLett.108.190602} {\bibfield  {journal} {\bibinfo  {journal} {Phys. Rev. Lett.}\ }\textbf {\bibinfo {volume} {108}},\ \bibinfo {pages} {190602} (\bibinfo {year} {2012})}\BibitemShut {NoStop}%
\bibitem [{\citenamefont {Zwanzig}(2001)}]{zwanzig2001}%
  \BibitemOpen
  \bibfield  {author} {\bibinfo {author} {\bibfnamefont {Robert}\ \bibnamefont {Zwanzig}},\ }\href@noop {} {\emph {\bibinfo {title} {Nonequilibrium Statistical Mechanics}}}\ (\bibinfo  {publisher} {Oxford University Press},\ \bibinfo {year} {2001})\BibitemShut {NoStop}%
\bibitem [{\citenamefont {Peliti}\ and\ \citenamefont {Pigolotti}(2021)}]{peliti2021}%
  \BibitemOpen
  \bibfield  {author} {\bibinfo {author} {\bibfnamefont {Luca}\ \bibnamefont {Peliti}}\ and\ \bibinfo {author} {\bibfnamefont {Simone}\ \bibnamefont {Pigolotti}},\ }\href@noop {} {\emph {\bibinfo {title} {Stochastic Thermodynamics: An Introduction}}}\ (\bibinfo  {publisher} {Princeton University Press},\ \bibinfo {year} {2021})\BibitemShut {NoStop}%
\bibitem [{\citenamefont {Blaber}\ and\ \citenamefont {Sivak}(2023)}]{sivak2023}%
  \BibitemOpen
  \bibfield  {author} {\bibinfo {author} {\bibfnamefont {Steven}\ \bibnamefont {Blaber}}\ and\ \bibinfo {author} {\bibfnamefont {David~A}\ \bibnamefont {Sivak}},\ }\bibfield  {title} {\enquote {\bibinfo {title} {Optimal control in stochastic thermodynamics},}\ }\href@noop {} {\bibfield  {journal} {\bibinfo  {journal} {Journal of Physics Communications}\ }\textbf {\bibinfo {volume} {7}},\ \bibinfo {pages} {033001} (\bibinfo {year} {2023})}\BibitemShut {NoStop}%
\bibitem [{\citenamefont {Esposito}\ \emph {et~al.}(2010)\citenamefont {Esposito}, \citenamefont {Kawai}, \citenamefont {Lindenberg},\ and\ \citenamefont {den Broeck}}]{Esposito_2010}%
  \BibitemOpen
  \bibfield  {author} {\bibinfo {author} {\bibfnamefont {M.}~\bibnamefont {Esposito}}, \bibinfo {author} {\bibfnamefont {R.}~\bibnamefont {Kawai}}, \bibinfo {author} {\bibfnamefont {K.}~\bibnamefont {Lindenberg}}, \ and\ \bibinfo {author} {\bibfnamefont {C.~Van}\ \bibnamefont {den Broeck}},\ }\bibfield  {title} {\enquote {\bibinfo {title} {Finite-time thermodynamics for a single-level quantum dot},}\ }\href {\doibase 10.1209/0295-5075/89/20003} {\bibfield  {journal} {\bibinfo  {journal} {Europhysics Letters}\ }\textbf {\bibinfo {volume} {89}},\ \bibinfo {pages} {20003} (\bibinfo {year} {2010})}\BibitemShut {NoStop}%
\bibitem [{\citenamefont {Diana}\ \emph {et~al.}(2013)\citenamefont {Diana}, \citenamefont {Bagci},\ and\ \citenamefont {Esposito}}]{Esposito2013}%
  \BibitemOpen
  \bibfield  {author} {\bibinfo {author} {\bibfnamefont {Giovanni}\ \bibnamefont {Diana}}, \bibinfo {author} {\bibfnamefont {G.~Baris}\ \bibnamefont {Bagci}}, \ and\ \bibinfo {author} {\bibfnamefont {Massimiliano}\ \bibnamefont {Esposito}},\ }\bibfield  {title} {\enquote {\bibinfo {title} {Finite-time erasing of information stored in fermionic bits},}\ }\href {\doibase 10.1103/PhysRevE.87.012111} {\bibfield  {journal} {\bibinfo  {journal} {Phys. Rev. E}\ }\textbf {\bibinfo {volume} {87}},\ \bibinfo {pages} {012111} (\bibinfo {year} {2013})}\BibitemShut {NoStop}%
\bibitem [{\citenamefont {Zulkowski}\ \emph {et~al.}(2013)\citenamefont {Zulkowski}, \citenamefont {Sivak},\ and\ \citenamefont {DeWeese}}]{zulkowski2013}%
  \BibitemOpen
  \bibfield  {author} {\bibinfo {author} {\bibfnamefont {Patrick~R}\ \bibnamefont {Zulkowski}}, \bibinfo {author} {\bibfnamefont {David~A}\ \bibnamefont {Sivak}}, \ and\ \bibinfo {author} {\bibfnamefont {Michael~R}\ \bibnamefont {DeWeese}},\ }\bibfield  {title} {\enquote {\bibinfo {title} {Optimal control of transitions between nonequilibrium steady states},}\ }\href@noop {} {\bibfield  {journal} {\bibinfo  {journal} {PloS one}\ }\textbf {\bibinfo {volume} {8}},\ \bibinfo {pages} {e82754} (\bibinfo {year} {2013})}\BibitemShut {NoStop}%
\bibitem [{\citenamefont {Lan}\ \emph {et~al.}(2012)\citenamefont {Lan}, \citenamefont {Sartori}, \citenamefont {Neumann}, \citenamefont {Sourjik},\ and\ \citenamefont {Tu}}]{tu2012}%
  \BibitemOpen
  \bibfield  {author} {\bibinfo {author} {\bibfnamefont {Ganhui}\ \bibnamefont {Lan}}, \bibinfo {author} {\bibfnamefont {Pablo}\ \bibnamefont {Sartori}}, \bibinfo {author} {\bibfnamefont {Silke}\ \bibnamefont {Neumann}}, \bibinfo {author} {\bibfnamefont {Victor}\ \bibnamefont {Sourjik}}, \ and\ \bibinfo {author} {\bibfnamefont {Yuhai}\ \bibnamefont {Tu}},\ }\bibfield  {title} {\enquote {\bibinfo {title} {The energy--speed--accuracy trade-off in sensory adaptation},}\ }\href@noop {} {\bibfield  {journal} {\bibinfo  {journal} {Nature physics}\ }\textbf {\bibinfo {volume} {8}},\ \bibinfo {pages} {422--428} (\bibinfo {year} {2012})}\BibitemShut {NoStop}%
\bibitem [{\citenamefont {Schmiedl}\ and\ \citenamefont {Seifert}(2007)}]{seifert2007}%
  \BibitemOpen
  \bibfield  {author} {\bibinfo {author} {\bibfnamefont {Tim}\ \bibnamefont {Schmiedl}}\ and\ \bibinfo {author} {\bibfnamefont {Udo}\ \bibnamefont {Seifert}},\ }\bibfield  {title} {\enquote {\bibinfo {title} {Optimal finite-time processes in stochastic thermodynamics},}\ }\href {\doibase 10.1103/PhysRevLett.98.108301} {\bibfield  {journal} {\bibinfo  {journal} {Phys. Rev. Lett.}\ }\textbf {\bibinfo {volume} {98}},\ \bibinfo {pages} {108301} (\bibinfo {year} {2007})}\BibitemShut {NoStop}%
\bibitem [{\citenamefont {Malaguti}\ and\ \citenamefont {ten Wolde}(2021)}]{malaguti2021}%
  \BibitemOpen
  \bibfield  {author} {\bibinfo {author} {\bibfnamefont {Giulia}\ \bibnamefont {Malaguti}}\ and\ \bibinfo {author} {\bibfnamefont {Pieter~Rein}\ \bibnamefont {ten Wolde}},\ }\bibfield  {title} {\enquote {\bibinfo {title} {Theory for the optimal detection of time-varying signals in cellular sensing systems},}\ }\href {\doibase 10.7554/eLife.62574} {\bibfield  {journal} {\bibinfo  {journal} {eLife}\ }\textbf {\bibinfo {volume} {10}},\ \bibinfo {pages} {e62574} (\bibinfo {year} {2021})}\BibitemShut {NoStop}%
\bibitem [{\citenamefont {Landauer}(1961)}]{landauer61}%
  \BibitemOpen
  \bibfield  {author} {\bibinfo {author} {\bibfnamefont {R.}~\bibnamefont {Landauer}},\ }\bibfield  {title} {\enquote {\bibinfo {title} {Irreversibility and heat generation in the computing process},}\ }\href {\doibase 10.1147/rd.53.0183} {\bibfield  {journal} {\bibinfo  {journal} {IBM Journal of Research and Development}\ }\textbf {\bibinfo {volume} {5}},\ \bibinfo {pages} {183--191} (\bibinfo {year} {1961})}\BibitemShut {NoStop}%
\bibitem [{\citenamefont {Theis}\ and\ \citenamefont {Wong}(2017)}]{wong17}%
  \BibitemOpen
  \bibfield  {author} {\bibinfo {author} {\bibfnamefont {Thomas~N.}\ \bibnamefont {Theis}}\ and\ \bibinfo {author} {\bibfnamefont {H.-S.~Philip}\ \bibnamefont {Wong}},\ }\bibfield  {title} {\enquote {\bibinfo {title} {The end of moore's law: A new beginning for information technology},}\ }\href {\doibase 10.1109/MCSE.2017.29} {\bibfield  {journal} {\bibinfo  {journal} {Computing in Science \& Engineering}\ }\textbf {\bibinfo {volume} {19}},\ \bibinfo {pages} {41--50} (\bibinfo {year} {2017})}\BibitemShut {NoStop}%
\bibitem [{\citenamefont {Szilard}(1929)}]{szilard29}%
  \BibitemOpen
  \bibfield  {author} {\bibinfo {author} {\bibfnamefont {Leo}\ \bibnamefont {Szilard}},\ }\bibfield  {title} {\enquote {\bibinfo {title} {{\"U}ber die {E}ntropieverminderung in einem thermodynamischen {S}ystem bei {E}ingriffen intelligenter {W}esen},}\ }\href@noop {} {\bibfield  {journal} {\bibinfo  {journal} {Zeitschrift f{\"u}r Physik}\ }\textbf {\bibinfo {volume} {53}},\ \bibinfo {pages} {840--856} (\bibinfo {year} {1929})}\BibitemShut {NoStop}%
\bibitem [{\citenamefont {Bennett}(1982)}]{Bennett82}%
  \BibitemOpen
  \bibfield  {author} {\bibinfo {author} {\bibfnamefont {Charles~H.}\ \bibnamefont {Bennett}},\ }\bibfield  {title} {\enquote {\bibinfo {title} {The thermodynamics of computation---a review},}\ }\href {\doibase 10.1007/BF02084158} {\bibfield  {journal} {\bibinfo  {journal} {International Journal of Theoretical Physics}\ }\textbf {\bibinfo {volume} {21}},\ \bibinfo {pages} {905--940} (\bibinfo {year} {1982})}\BibitemShut {NoStop}%
\bibitem [{\citenamefont {Sagawa}\ and\ \citenamefont {Ueda}(2009)}]{ueda09}%
  \BibitemOpen
  \bibfield  {author} {\bibinfo {author} {\bibfnamefont {Takahiro}\ \bibnamefont {Sagawa}}\ and\ \bibinfo {author} {\bibfnamefont {Masahito}\ \bibnamefont {Ueda}},\ }\bibfield  {title} {\enquote {\bibinfo {title} {Minimal energy cost for thermodynamic information processing: Measurement and information erasure},}\ }\href {\doibase 10.1103/PhysRevLett.102.250602} {\bibfield  {journal} {\bibinfo  {journal} {Phys. Rev. Lett.}\ }\textbf {\bibinfo {volume} {102}},\ \bibinfo {pages} {250602} (\bibinfo {year} {2009})}\BibitemShut {NoStop}%
\bibitem [{\citenamefont {Ouldridge}\ \emph {et~al.}(2019)\citenamefont {Ouldridge}, \citenamefont {Brittain},\ and\ \citenamefont {ten Wolde}}]{tenwolde19}%
  \BibitemOpen
  \bibfield  {author} {\bibinfo {author} {\bibfnamefont {Thomas~E.}\ \bibnamefont {Ouldridge}}, \bibinfo {author} {\bibfnamefont {Rory~A.}\ \bibnamefont {Brittain}}, \ and\ \bibinfo {author} {\bibfnamefont {Pieter~Rein}\ \bibnamefont {ten Wolde}},\ }\bibfield  {title} {\enquote {\bibinfo {title} {The power of being explicit: Demystifying work, heat, and free energy in the physics of computation},}\ }in\ \href@noop {} {\emph {\bibinfo {booktitle} {The Energetics of Computing in Life and Machines}}},\ \bibinfo {editor} {edited by\ \bibinfo {editor} {\bibfnamefont {David~H.}\ \bibnamefont {Wolpert}}, \bibinfo {editor} {\bibfnamefont {Chris}\ \bibnamefont {Kempes}}, \bibinfo {editor} {\bibfnamefont {Peter~F.}\ \bibnamefont {Stadler}}, \ and\ \bibinfo {editor} {\bibfnamefont {Joshua~A.}\ \bibnamefont {Grochow}}}\ (\bibinfo  {publisher} {SFI Press},\ \bibinfo {address} {Santa Fe},\ \bibinfo {year} {2019})\ Chap.~\bibinfo {chapter} {12}, pp.\ \bibinfo {pages} {307--351}\BibitemShut {NoStop}%
\bibitem [{\citenamefont {Mulder}\ \emph {et~al.}(2023)\citenamefont {Mulder}, \citenamefont {ten Wolde},\ and\ \citenamefont {Ouldridge}}]{mulder2023}%
  \BibitemOpen
  \bibfield  {author} {\bibinfo {author} {\bibfnamefont {Daan}\ \bibnamefont {Mulder}}, \bibinfo {author} {\bibfnamefont {Pieter~Rein}\ \bibnamefont {ten Wolde}}, \ and\ \bibinfo {author} {\bibfnamefont {Thomas~E.}\ \bibnamefont {Ouldridge}},\ }\bibfield  {title} {\enquote {\bibinfo {title} {Exploiting bias in optimal finite-time copying protocols},}\ }\href@noop {} {\  (\bibinfo {year} {2023})},\ \Eprint {http://arxiv.org/abs/2312.14682} {arXiv:2312.14682 [cond-mat.stat-mech]} \BibitemShut {NoStop}%
\bibitem [{\citenamefont {Koski}\ \emph {et~al.}(2014)\citenamefont {Koski}, \citenamefont {Maisi}, \citenamefont {Pekola},\ and\ \citenamefont {Averin}}]{Koski.2014tu}%
  \BibitemOpen
  \bibfield  {author} {\bibinfo {author} {\bibfnamefont {Jonne~V.}\ \bibnamefont {Koski}}, \bibinfo {author} {\bibfnamefont {Ville~F.}\ \bibnamefont {Maisi}}, \bibinfo {author} {\bibfnamefont {Jukka~P.}\ \bibnamefont {Pekola}}, \ and\ \bibinfo {author} {\bibfnamefont {Dmitri~V.}\ \bibnamefont {Averin}},\ }\bibfield  {title} {\enquote {\bibinfo {title} {{Experimental realization of a Szilard engine with a single electron}},}\ }\href {\doibase 10.1073/pnas.1406966111} {\bibfield  {journal} {\bibinfo  {journal} {Proceedings of the National Academy of Sciences}\ }\textbf {\bibinfo {volume} {111}},\ \bibinfo {pages} {13786--13789} (\bibinfo {year} {2014})}\BibitemShut {NoStop}%
\bibitem [{\citenamefont {Jun}\ \emph {et~al.}(2014)\citenamefont {Jun}, \citenamefont {Gavrilov},\ and\ \citenamefont {Bechhoefer}}]{Jun.2014qmn}%
  \BibitemOpen
  \bibfield  {author} {\bibinfo {author} {\bibfnamefont {Yonggun}\ \bibnamefont {Jun}}, \bibinfo {author} {\bibfnamefont {Momčilo}\ \bibnamefont {Gavrilov}}, \ and\ \bibinfo {author} {\bibfnamefont {John}\ \bibnamefont {Bechhoefer}},\ }\bibfield  {title} {{\selectlanguage {English}\enquote {\bibinfo {title} {{High-Precision Test of Landauer’s Principle in a Feedback Trap}},}\ }}\href {\doibase 10.1103/physrevlett.113.190601} {\bibfield  {journal} {\bibinfo  {journal} {Physical Review Letters}\ }\textbf {\bibinfo {volume} {113}},\ \bibinfo {pages} {190601} (\bibinfo {year} {2014})}\BibitemShut {NoStop}%
\bibitem [{\citenamefont {B\'{e}rut}\ \emph {et~al.}(2012)\citenamefont {B\'{e}rut}, \citenamefont {Arakelyan}, \citenamefont {Petrosyan}, \citenamefont {Ciliberto}, \citenamefont {Dillenschneider},\ and\ \citenamefont {Lutz}}]{Berut.2012}%
  \BibitemOpen
  \bibfield  {author} {\bibinfo {author} {\bibfnamefont {Antoine}\ \bibnamefont {B\'{e}rut}}, \bibinfo {author} {\bibfnamefont {Artak}\ \bibnamefont {Arakelyan}}, \bibinfo {author} {\bibfnamefont {Artyom}\ \bibnamefont {Petrosyan}}, \bibinfo {author} {\bibfnamefont {Sergio}\ \bibnamefont {Ciliberto}}, \bibinfo {author} {\bibfnamefont {Raoul}\ \bibnamefont {Dillenschneider}}, \ and\ \bibinfo {author} {\bibfnamefont {Eric}\ \bibnamefont {Lutz}},\ }\bibfield  {title} {\enquote {\bibinfo {title} {{Experimental verification of Landauer’s principle linking information and thermodynamics}},}\ }\href {\doibase 10.1038/nature10872} {\bibfield  {journal} {\bibinfo  {journal} {Nature}\ }\textbf {\bibinfo {volume} {483}},\ \bibinfo {pages} {187 -- 189} (\bibinfo {year} {2012})}\BibitemShut {NoStop}%
\bibitem [{\citenamefont {Proesmans}\ \emph {et~al.}(2020)\citenamefont {Proesmans}, \citenamefont {Ehrich},\ and\ \citenamefont {Bechhoefer}}]{bechhoefer20}%
  \BibitemOpen
  \bibfield  {author} {\bibinfo {author} {\bibfnamefont {Karel}\ \bibnamefont {Proesmans}}, \bibinfo {author} {\bibfnamefont {Jannik}\ \bibnamefont {Ehrich}}, \ and\ \bibinfo {author} {\bibfnamefont {John}\ \bibnamefont {Bechhoefer}},\ }\bibfield  {title} {\enquote {\bibinfo {title} {Finite-time {L}andauer principle},}\ }\href {\doibase 10.1103/PhysRevLett.125.100602} {\bibfield  {journal} {\bibinfo  {journal} {Phys. Rev. Lett.}\ }\textbf {\bibinfo {volume} {125}},\ \bibinfo {pages} {100602} (\bibinfo {year} {2020})}\BibitemShut {NoStop}%
\bibitem [{\citenamefont {Maes}(2017)}]{maes2017}%
  \BibitemOpen
  \bibfield  {author} {\bibinfo {author} {\bibfnamefont {Christian}\ \bibnamefont {Maes}},\ }\href@noop {} {\emph {\bibinfo {title} {Non-dissipative effects in nonequilibrium systems}}}\ (\bibinfo  {publisher} {Springer},\ \bibinfo {year} {2017})\BibitemShut {NoStop}%
\bibitem [{\citenamefont {Bauer}\ \emph {et~al.}(2014)\citenamefont {Bauer}, \citenamefont {Barato},\ and\ \citenamefont {Seifert}}]{Seifert_2014}%
  \BibitemOpen
  \bibfield  {author} {\bibinfo {author} {\bibfnamefont {Michael}\ \bibnamefont {Bauer}}, \bibinfo {author} {\bibfnamefont {Andre~C}\ \bibnamefont {Barato}}, \ and\ \bibinfo {author} {\bibfnamefont {Udo}\ \bibnamefont {Seifert}},\ }\bibfield  {title} {\enquote {\bibinfo {title} {Optimized finite-time information machine},}\ }\href {\doibase 10.1088/1742-5468/2014/09/P09010} {\bibfield  {journal} {\bibinfo  {journal} {Journal of Statistical Mechanics: Theory and Experiment}\ }\textbf {\bibinfo {volume} {2014}},\ \bibinfo {pages} {P09010} (\bibinfo {year} {2014})}\BibitemShut {NoStop}%
\bibitem [{\citenamefont {Lee}\ \emph {et~al.}(2022)\citenamefont {Lee}, \citenamefont {Lee}, \citenamefont {Kwon},\ and\ \citenamefont {Park}}]{park22}%
  \BibitemOpen
  \bibfield  {author} {\bibinfo {author} {\bibfnamefont {Jae~Sung}\ \bibnamefont {Lee}}, \bibinfo {author} {\bibfnamefont {Sangyun}\ \bibnamefont {Lee}}, \bibinfo {author} {\bibfnamefont {Hyukjoon}\ \bibnamefont {Kwon}}, \ and\ \bibinfo {author} {\bibfnamefont {Hyunggyu}\ \bibnamefont {Park}},\ }\bibfield  {title} {\enquote {\bibinfo {title} {Speed limit for a highly irreversible process and tight finite-time {L}andauer's bound},}\ }\href {\doibase 10.1103/PhysRevLett.129.120603} {\bibfield  {journal} {\bibinfo  {journal} {Phys. Rev. Lett.}\ }\textbf {\bibinfo {volume} {129}},\ \bibinfo {pages} {120603} (\bibinfo {year} {2022})}\BibitemShut {NoStop}%
\bibitem [{\citenamefont {Dechant}(2022)}]{dechant2022}%
  \BibitemOpen
  \bibfield  {author} {\bibinfo {author} {\bibfnamefont {Andreas}\ \bibnamefont {Dechant}},\ }\bibfield  {title} {\enquote {\bibinfo {title} {Minimum entropy production, detailed balance and {W}asserstein distance for continuous-time {M}arkov processes},}\ }\href {\doibase 10.1088/1751-8121/ac4ac0} {\bibfield  {journal} {\bibinfo  {journal} {Journal of Physics A: Mathematical and Theoretical}\ }\textbf {\bibinfo {volume} {55}},\ \bibinfo {pages} {094001} (\bibinfo {year} {2022})}\BibitemShut {NoStop}%
\bibitem [{\citenamefont {Remlein}\ and\ \citenamefont {Seifert}(2021)}]{Remlein.2021}%
  \BibitemOpen
  \bibfield  {author} {\bibinfo {author} {\bibfnamefont {Benedikt}\ \bibnamefont {Remlein}}\ and\ \bibinfo {author} {\bibfnamefont {Udo}\ \bibnamefont {Seifert}},\ }\bibfield  {title} {\enquote {\bibinfo {title} {Optimality of nonconservative driving for finite-time processes with discrete states},}\ }\href {\doibase 10.1103/PhysRevE.103.L050105} {\bibfield  {journal} {\bibinfo  {journal} {Phys. Rev. E}\ }\textbf {\bibinfo {volume} {103}},\ \bibinfo {pages} {L050105} (\bibinfo {year} {2021})}\BibitemShut {NoStop}%
\bibitem [{\citenamefont {Zulkowski}\ and\ \citenamefont {DeWeese}(2014)}]{deweese14}%
  \BibitemOpen
  \bibfield  {author} {\bibinfo {author} {\bibfnamefont {Patrick~R.}\ \bibnamefont {Zulkowski}}\ and\ \bibinfo {author} {\bibfnamefont {Michael~R.}\ \bibnamefont {DeWeese}},\ }\bibfield  {title} {\enquote {\bibinfo {title} {Optimal finite-time erasure of a classical bit},}\ }\href {\doibase 10.1103/PhysRevE.89.052140} {\bibfield  {journal} {\bibinfo  {journal} {Phys. Rev. E}\ }\textbf {\bibinfo {volume} {89}},\ \bibinfo {pages} {052140} (\bibinfo {year} {2014})}\BibitemShut {NoStop}%
\bibitem [{\citenamefont {Van~Vu}\ and\ \citenamefont {Saito}(2023)}]{Vu.2023}%
  \BibitemOpen
  \bibfield  {author} {\bibinfo {author} {\bibfnamefont {Tan}\ \bibnamefont {Van~Vu}}\ and\ \bibinfo {author} {\bibfnamefont {Keiji}\ \bibnamefont {Saito}},\ }\bibfield  {title} {\enquote {\bibinfo {title} {{Thermodynamic Unification of Optimal Transport: Thermodynamic Uncertainty Relation, Minimum Dissipation, and Thermodynamic Speed Limits}},}\ }\href {\doibase 10.1103/physrevx.13.011013} {\bibfield  {journal} {\bibinfo  {journal} {Physical Review X}\ }\textbf {\bibinfo {volume} {13}},\ \bibinfo {pages} {011013} (\bibinfo {year} {2023})},\ \Eprint {http://arxiv.org/abs/2206.02684} {2206.02684} \BibitemShut {NoStop}%
\bibitem [{\citenamefont {Shiraishi}\ \emph {et~al.}(2018)\citenamefont {Shiraishi}, \citenamefont {Funo},\ and\ \citenamefont {Saito}}]{shiraishi18}%
  \BibitemOpen
  \bibfield  {author} {\bibinfo {author} {\bibfnamefont {Naoto}\ \bibnamefont {Shiraishi}}, \bibinfo {author} {\bibfnamefont {Ken}\ \bibnamefont {Funo}}, \ and\ \bibinfo {author} {\bibfnamefont {Keiji}\ \bibnamefont {Saito}},\ }\bibfield  {title} {\enquote {\bibinfo {title} {Speed limit for classical stochastic processes},}\ }\href {\doibase 10.1103/PhysRevLett.121.070601} {\bibfield  {journal} {\bibinfo  {journal} {Phys. Rev. Lett.}\ }\textbf {\bibinfo {volume} {121}},\ \bibinfo {pages} {070601} (\bibinfo {year} {2018})}\BibitemShut {NoStop}%
\bibitem [{SI()}]{SI}%
  \BibitemOpen
  \href@noop {} {}\bibinfo {note} {See Supplemental Material for derivations.}\BibitemShut {Stop}%
\bibitem [{\citenamefont {Strasberg}\ \emph {et~al.}(2017)\citenamefont {Strasberg}, \citenamefont {Schaller}, \citenamefont {Brandes},\ and\ \citenamefont {Esposito}}]{esposito2017}%
  \BibitemOpen
  \bibfield  {author} {\bibinfo {author} {\bibfnamefont {Philipp}\ \bibnamefont {Strasberg}}, \bibinfo {author} {\bibfnamefont {Gernot}\ \bibnamefont {Schaller}}, \bibinfo {author} {\bibfnamefont {Tobias}\ \bibnamefont {Brandes}}, \ and\ \bibinfo {author} {\bibfnamefont {Massimiliano}\ \bibnamefont {Esposito}},\ }\bibfield  {title} {\enquote {\bibinfo {title} {Quantum and information thermodynamics: A unifying framework based on repeated interactions},}\ }\href {\doibase 10.1103/PhysRevX.7.021003} {\bibfield  {journal} {\bibinfo  {journal} {Phys. Rev. X}\ }\textbf {\bibinfo {volume} {7}},\ \bibinfo {pages} {021003} (\bibinfo {year} {2017})}\BibitemShut {NoStop}%
\bibitem [{\citenamefont {Maes}\ \emph {et~al.}(2008)\citenamefont {Maes}, \citenamefont {Netocn{\`y}},\ and\ \citenamefont {Wynants}}]{maes2008}%
  \BibitemOpen
  \bibfield  {author} {\bibinfo {author} {\bibfnamefont {Christian}\ \bibnamefont {Maes}}, \bibinfo {author} {\bibfnamefont {Karel}\ \bibnamefont {Netocn{\`y}}}, \ and\ \bibinfo {author} {\bibfnamefont {Bram}\ \bibnamefont {Wynants}},\ }\bibfield  {title} {\enquote {\bibinfo {title} {On and beyond entropy production: the case of {M}arkov jump processes},}\ }in\ \href@noop {} {\emph {\bibinfo {booktitle} {Markov Proc. Rel. Fields}}},\ Vol.~\bibinfo {volume} {14}\ (\bibinfo {year} {2008})\ pp.\ \bibinfo {pages} {445--464}\BibitemShut {NoStop}%
\bibitem [{\citenamefont {Van~Vu}\ and\ \citenamefont {Hasegawa}(2021)}]{hasegawa2021}%
  \BibitemOpen
  \bibfield  {author} {\bibinfo {author} {\bibfnamefont {Tan}\ \bibnamefont {Van~Vu}}\ and\ \bibinfo {author} {\bibfnamefont {Yoshihiko}\ \bibnamefont {Hasegawa}},\ }\bibfield  {title} {\enquote {\bibinfo {title} {Geometrical bounds of the irreversibility in markovian systems},}\ }\href {\doibase 10.1103/PhysRevLett.126.010601} {\bibfield  {journal} {\bibinfo  {journal} {Phys. Rev. Lett.}\ }\textbf {\bibinfo {volume} {126}},\ \bibinfo {pages} {010601} (\bibinfo {year} {2021})}\BibitemShut {NoStop}%
\bibitem [{\citenamefont {Aurell}\ \emph {et~al.}(2011)\citenamefont {Aurell}, \citenamefont {Mej\'{\i}a-Monasterio},\ and\ \citenamefont {Muratore-Ginanneschi}}]{aurell11}%
  \BibitemOpen
  \bibfield  {author} {\bibinfo {author} {\bibfnamefont {Erik}\ \bibnamefont {Aurell}}, \bibinfo {author} {\bibfnamefont {Carlos}\ \bibnamefont {Mej\'{\i}a-Monasterio}}, \ and\ \bibinfo {author} {\bibfnamefont {Paolo}\ \bibnamefont {Muratore-Ginanneschi}},\ }\bibfield  {title} {\enquote {\bibinfo {title} {Optimal protocols and optimal transport in stochastic thermodynamics},}\ }\href {\doibase 10.1103/PhysRevLett.106.250601} {\bibfield  {journal} {\bibinfo  {journal} {Phys. Rev. Lett.}\ }\textbf {\bibinfo {volume} {106}},\ \bibinfo {pages} {250601} (\bibinfo {year} {2011})}\BibitemShut {NoStop}%
\bibitem [{\citenamefont {Nakazato}\ and\ \citenamefont {Ito}(2021)}]{ito21}%
  \BibitemOpen
  \bibfield  {author} {\bibinfo {author} {\bibfnamefont {Muka}\ \bibnamefont {Nakazato}}\ and\ \bibinfo {author} {\bibfnamefont {Sosuke}\ \bibnamefont {Ito}},\ }\bibfield  {title} {\enquote {\bibinfo {title} {Geometrical aspects of entropy production in stochastic thermodynamics based on {W}asserstein distance},}\ }\href {\doibase 10.1103/PhysRevResearch.3.043093} {\bibfield  {journal} {\bibinfo  {journal} {Phys. Rev. Res.}\ }\textbf {\bibinfo {volume} {3}},\ \bibinfo {pages} {043093} (\bibinfo {year} {2021})}\BibitemShut {NoStop}%
\bibitem [{\citenamefont {Nagayama}\ \emph {et~al.}(2024)\citenamefont {Nagayama}, \citenamefont {Yoshimura},\ and\ \citenamefont {Ito}}]{ito2024}%
  \BibitemOpen
  \bibfield  {author} {\bibinfo {author} {\bibfnamefont {Ryuna}\ \bibnamefont {Nagayama}}, \bibinfo {author} {\bibfnamefont {Kohei}\ \bibnamefont {Yoshimura}}, \ and\ \bibinfo {author} {\bibfnamefont {Sosuke}\ \bibnamefont {Ito}},\ }\href {https://arxiv.org/abs/2412.20690} {\enquote {\bibinfo {title} {Infinite variety of thermodynamic speed limits with general activities},}\ } (\bibinfo {year} {2024}),\ \Eprint {http://arxiv.org/abs/2412.20690} {arXiv:2412.20690 [cond-mat.stat-mech]} \BibitemShut {NoStop}%
\bibitem [{\citenamefont {Sivak}\ and\ \citenamefont {Crooks}(2016)}]{sivak2016}%
  \BibitemOpen
  \bibfield  {author} {\bibinfo {author} {\bibfnamefont {David~A.}\ \bibnamefont {Sivak}}\ and\ \bibinfo {author} {\bibfnamefont {Gavin~E.}\ \bibnamefont {Crooks}},\ }\bibfield  {title} {\enquote {\bibinfo {title} {Thermodynamic geometry of minimum-dissipation driven barrier crossing},}\ }\href {\doibase 10.1103/PhysRevE.94.052106} {\bibfield  {journal} {\bibinfo  {journal} {Phys. Rev. E}\ }\textbf {\bibinfo {volume} {94}},\ \bibinfo {pages} {052106} (\bibinfo {year} {2016})}\BibitemShut {NoStop}%
\end{thebibliography}
\end{document}